\documentclass[prd,aps,nofootinbib,preprintnumbers,showpacs,floatfix]{revtex4}
\usepackage{amsmath,amssymb}
\usepackage{epsfig}
\usepackage{graphicx}
\usepackage{subfigure}

\begin{document}
\title{Preheating after Small-Field Inflation}
\date{\today}
\author{Philippe Brax$^{1}$, Jean-Fran\c{c}ois Dufaux$^{2}$, Sophie Mariadassou$^{1}$}
\affiliation{$^{1}$Institut de Physique Th\'eorique, CEA, IPhT, CNRS, URA 2306, F-91191Gif/Yvette Cedex, France\\
$^{2}$APC, UMR 7164 (CNRS - Universit\'e Paris 7), 10 rue Alice Domon et L\'eonie Duquet, 75205 Paris Cedex 13, France}
\preprint{}

\def\O{\mathcal{O}}
\def\lesssim{\mathrel{\hbox{\rlap{\hbox{\lower4pt\hbox{$\sim$}}}\hbox{$<$}}}}
\def\gtrsim{\mathrel{\hbox{\rlap{\hbox{\lower4pt\hbox{$\sim$}}}\hbox{$>$}}}}
\def\Mp{M_{\mathrm{Pl}}}
\def\mp{m_{\mathrm{Pl}}}
\def\be{\begin{equation}}
\def\ee{\end{equation}}
\def\bea{\begin{eqnarray}}
\def\eea{\end{eqnarray}}
\newcommand{\picdir}[1]{../Figs/#1}
\newcommand{\ba}{\begin{array}}
\newcommand{\ea}{\end{array}}
\newcommand{\nn}{\nonumber \\}
\newcommand{\lag}{{\mathcal L}}
\newcommand{\mn}{{\mu\nu}}
\newcommand{\ab}{{\alpha\beta}}
\newcommand{\hI}{\hspace{1cm}}
\newcommand{\hVII}{\hspace{.7cm}}
\newcommand{\hV}{\hspace{.5cm}}
\newcommand{\hu}{{\mathcal H}}
\newcommand{\bq}{{\bf q}}
\newcommand{\bk}{{\bf k}}
\newcommand{\bx}{{\bf x}}

\begin{abstract}
Whereas preheating after  chaotic and hybrid inflation models has been abundantly studied in the literature,
preheating in small field inflation models, where the curvature of the inflaton potential is negative
during inflation, remains less explored. In these models, a tachyonic instability at the end of inflation leads to a succession of exponentially large increases and \emph{decreases} of the inflaton fluctuations as the inflaton condensate oscillates around the minimum of its potential. The net effect is  a competition between low-momentum modes which grow and decrease significantly, and modes with higher momenta which grow less but also decrease less. We develop an analytical
description of this process, which is analogous to the quantum mechanical problem of tunneling through a volcano-shaped
potential. Depending on the parameters, preheating may be so efficient that it completes in less than one oscillation
of the inflaton condensate. Preheating after small field inflation may also be followed by a long matter-dominated stage before the universe thermalizes, depending on the energy scale of inflation and the details of the inflaton interactions.
Finally, another feature of these models is that the spectrum of the inflaton fluctuations at the end of preheating may be peaked around the Hubble scale.
In fact, because preheating starts when the second slow-roll parameter $|\eta|$ becomes of order unity while the first slow-roll parameter $\epsilon$ is still much
smaller than one, the universe is still inflating during preheating and the modes amplified by the initial tachyonic instability leave the Hubble radius. This
may lead to an abundant production of primordial black holes and gravitational waves with frequencies today which  are naturally small enough to fall into the range
accessible by high-sensitivity interferometric experiments.
\end{abstract}

\maketitle

%%%%%%%%%%%%%%%%%%%%%%%%%%%%%%%%%%%%%%%%%%%%%%%%%%%%%%%%%%%%%%%%%%%%%%%%%

\section{Introduction}

Simple inflationary models can be broadly but conveniently classified into three main categories, see
e.g.~\cite{FieldRev,StringRev} for reviews. One category comprises hybrid inflation models~\cite{hybrid}, where the inflaton is responsible for the slow-roll dynamics while another scalar field is responsible for the end of inflation. Interesting aspects of these models include the possible production of topological defects at the end of inflation and possible realisations in supergravity as F-term or D-term inflation~\cite{FieldRev} and in string theory in the framework of brane inflation~\cite{StringRev}. Another category comprising chaotic inflation~\cite{chaotic} includes the large field inflation models. They have the simplest field theoretic origin as a single massive scalar field is enough to generate inflation. Another relevant feature of these models is the large, super-Planckian value of the inflaton and therefore the possibility to generate a significant amount of primordial gravity waves during inflation. Unfortunately,
such large values of the inflaton may render these models more difficult to realize in supergravity or string theory. In
this respect, the category of small field or hilltop inflation models~\cite{hilltop}, which includes for instance the original new inflation model~\cite{new}, may be more natural. In these models, inflation occurs near a maximum or an inflection point of the potential, where the curvature is negative. This makes the slow-roll conditions somewhat easier
to achieve and leads to cosmological perturbations with a negative spectral index as favoured by observations. Another interesting aspect of these models is that inflation can occur at very low energy scales, while still generating
an acceptable spectrum of primordial density perturbations~\cite{lowscale}. These models offer also string theoretic realisations, for instance in the form of racetrack models~\cite{StringRev} where the imaginary part of a K\"ahler modulus acts as the inflaton. Another consequence of small field inflation follows from the Lyth bound on the inflaton excursion~\cite{lythbound}, which implies that gravitational waves generated during inflation are highly suppressed in this setting.

At the end of inflation, the inflaton condensate must decay and reheat the universe. It has become clear during the last
twenty years or so~\cite{traschen} that, in most models, reheating starts with an explosive and non-perturbative
production of large, non-thermal fluctuations of the inflaton and other bosonic fields coupled to it, in the process
of preheating~\cite{KLS}. The subsequent dynamics are characterized by a highly non-linear and turbulent-like evolution, before the system eventually settles into thermal equilibrium. Preheating may have many interesting consequences in cosmology, like the production of stochastic backgrounds of gravitational waves~\cite{GWcha, GWhyb, GWpre, GWflat, GWvec}, primordial black holes~\cite{malik, bassett, kudoh1, kudoh2}, non-gaussian curvature perturbations~\cite{rajantie, bond}
or primordial magnetic fields~\cite{magnetic}. Preheating after chaotic and hybrid inflation has been abundantly studied
in the literature, both analytically and with numerical lattice simulations~\cite{KTlatt, latticeeasy, defrost}. In models of chaotic inflation, preheating occurs typically via broad parametric resonance~\cite{KLS}, where fields coupled to the inflaton are produced non-adiabatically as the inflaton oscillates around the minimum of its potential, or via tachyonic resonance~\cite{tachres} in the presence of trilinear or non-renormalizable bosonic interactions for the inflaton. In models of hybrid inflation, field fluctuations are amplified by a tachyonic (spinodal) instability~\cite{tachyonic}
as the fields roll towards the true minimum of the potential at the end of inflation.

By contrast, preheating after small field inflation remains less explored. On the analytical side, the problem is complicated by the fact that closed-form solutions for the evolution of the homogeneous background during preheating
are in general not available, while standard approximations like WKB for the evolution of the perturbations are not applicable in this context. On the numerical side, lattice simulations of preheating in these models must cover scales
that range from the Hubble rate at the end of inflation up to the inflaton mass at the minimum of the potential, which typically differ by several orders of magnitude. Preheating after new inflation was studied in \cite{PreNew}\footnote{See \cite{traschen} for earlier work.}, where it was described as a combination of both tachyonic amplification and non-adiabatic resonance. In this sense, it is somehow intermediate between preheating after chaotic inflation and preheating after hybrid inflation. Some aspects of preheating in this model where also studied numerically in \cite{kudoh2} and in the context of K\"ahler moduli / Roulette inflation in \cite{PreRoul}. The combination of tachyonic and non-adiabatic effects occurs also in preheating via trilinear or non-renormalizable interactions after chaotic inflation~\cite{tachres}. However, we will see that preheating after small field inflation occurs in a qualitatively different way.

\begin{figure}[htb]
\begin{center}
\includegraphics[width=13cm]{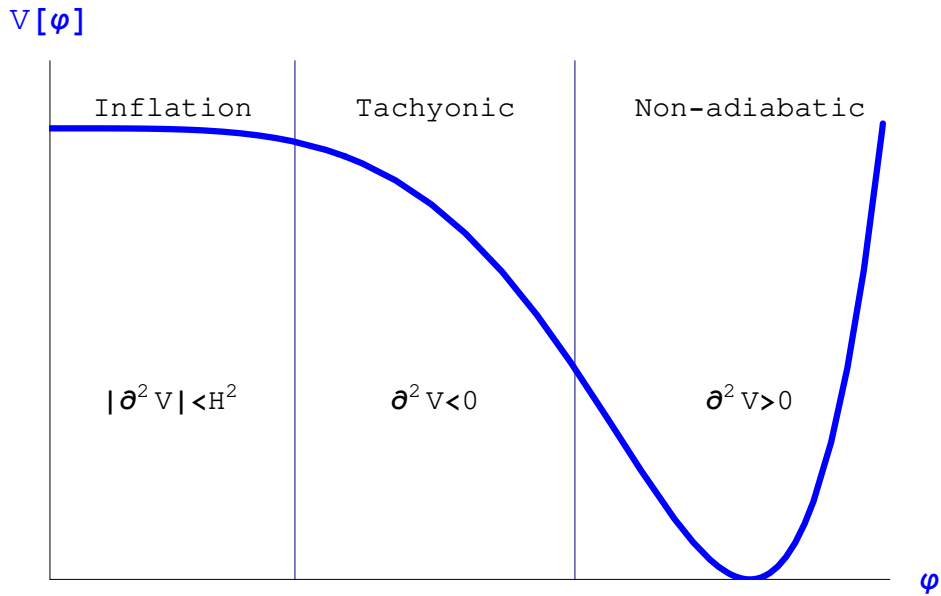}
\end{center}
\vspace*{-5mm}
\caption{Schematic form of the inflaton potential in a model of small field inflation.}
\label{smallpot}
\vspace*{-3mm}
\end{figure}

The main purpose of this paper is to develop the analytical understanding of preheating in the class of small field inflation models. This is a first necessary step before a further study of the dynamics with lattice simulations and of
the cosmological consequences of preheating in these models. To illustrate some of the qualitative features of
preheating after small field inflation, consider the inflaton potential shown in Fig.~\ref{smallpot}. Slow-roll inflation ends when the second slow-roll parameter $|\eta|$ becomes of order one, where the curvature of the potential is
negative and of the order of the Hubble rate squared. The inflaton then rolls down the potential, passes through an inflection point where the curvature vanishes and oscillates around the minimum with an amplitude controlled by Hubble friction. Before the inflaton reaches the inflection point, its fluctuations have a negative effective mass squared and modes with momentum $k^2 < -V''(\phi)$ are amplified by a tachyonic instability. This process may affect a very wide range
of scales, from the Hubble rate at the end of inflation up to the maximum of $|V''(\phi)|$, but it is much more
efficient for the low-momentum modes. The tachyonic amplification stops when the inflaton crosses the inflexion point
of the potential, oscillates around the minimum and goes back towards the tachyonic region. During this interval of
time, the mass of the inflaton varies very non-adiabatically with time. This can lead to a further growth of the perturbations, but we will see that this effect is typically negligible. However, the evolution of the modes during this time interval when the inflaton oscillates around the minimum of the potential has dramatic consequences for the fate of the fluctuations when the inflaton goes back into the region where the curvature is negative. Indeed, during this second tachyonic episode, the amplitude of the fluctuations starts to decrease exponentially with time. For the modes with sufficiently low momenta, this exponential decrease occurs during  the amount of time  the inflaton condensate climbs back along  its potential. If the inflaton climbed back up to the point from where it started, this effect would exactly compensate
the growth of these modes during the first tachyonic episode. By contrast, the modes with higher momentum, which were amplified at a slower rate during the first tachyonic episode, continue to be amplified during the second one. The whole process may then repeat itself several times as the inflaton oscillates around the minimum of the potential. The net
effect then follows from a competition between low-momentum modes which grow a lot but also decrease a lot and modes with higher momenta which grow less but also decrease less. As we will see, the analytical description of this process is very similar to the problem of tunneling through a volcano-shaped potential (two potential barriers separated by a deep potential well) in quantum mechanics. Alternatively, the initial tachyonic growth may be so efficient that preheating ends during the first tachyonic episode, depending on the parameters.

As we will see, another specificity of small field models is that the spectrum of the inflaton fluctuations amplified at the beginning of preheating is peaked
around the Hubble rate, because the modes with $k/a \lesssim H$ are amplified from the beginning of preheating when $V'' \sim -H^2$ while the modes with higher
momenta are amplified only later and at a slower rate. The situation is different for preheating after chaotic inflation, where the fluctuations amplified by
parametric resonance are typically peaked at scales $k / (aH) > q^{1/4} \gg 1$ with the initial resonance parameter $q \gg 1$ for preheating to be
efficient~\cite{KLS}. Similarly, for preheating after hybrid inflation, the typical scale of the fluctuations amplified by the tachyonic instability is set by
the curvature of the potential in the direction of the symmetry-breaking field and must be sub-Hubble to satisfy the so-called waterfall condition~\cite{hybrid}.
In fact, in models of small field inflation, the modes amplified by the initial tachyonic instability leave the Hubble radius, because preheating starts when
the the second slow-roll parameter $|\eta|$ becomes of order unity while the first slow-roll parameter $\epsilon$ is still much smaller than one, so that the
universe is still inflating at the beginning of preheating.  When preheating ends in less than one oscillation of the inflaton condensate, large density
perturbations of the Hubble size may lead to an abundant production of primordial black holes, see also~\cite{kudoh2}, which may put strong constraints on models of
small field inflation. This is in contrast with preheating after chaotic and hybrid inflation, where the field inhomogeneities are peaked at scales that are much shorter than the Hubble
radius~\cite{kudoh1,defrost}. Furthermore, large field fluctuations at the Hubble scale and the fact that small field inflation can occur at very low energy scales
may lead to the production of gravitational waves from preheating with frequencies today that are sufficiently small to fall into the range accessible by
high-sensitivity ground-based and even space-based interferometric experiments~\cite{GWpre}. A detailed study of these cosmological consequences of preheating after small field inflation will
appear elsewhere~\cite{paper2}.

The rest of the paper is organized as follows. In section \ref{SecBack}, we specify the models of small field inflation that we will consider and we describe
the inflationary dynamics and the subsequent evolution of the inflaton condensate. Section \ref{Sec1} is dedicated to the first stage of preheating, when
the inflaton condensate first rolls towards the minimum of the potential. In Section \ref{Sec2}, we study the second stage of preheating, when the inflaton
condensate oscillates around the minimum. Some of the results used in that Section are derived in the Appendix. In Section~\ref{SecPert}, we first consider
other possible non-perturbative decay channels for the condensate. We then discuss the perturbative decay of the inflaton after preheating and the resulting
reheat temperature. Finally, Section \ref{SecConclu} contains a summary of our results and directions for future work.

%%%%%%%%%%%%%%%%%%%%%%%%%%%%%%%%%%%%%%%%%%%%%%%%%%%%%%%%%%%%%%%%%%%%%%%%%%%

\section{Small Field Inflation and Background Evolution}
\label{SecBack}

We consider small field inflation models with a single inflaton field $\phi$. Inflation occurs near a maximum
or an inflection point of the potential, which can be chosen as the origin $\phi = 0$. Denoting by $M$ the energy
scale during inflation, the potential can be parametrized as
\be
\label{pot}
V(\phi) = M^4\,\left[1 - f\left(\frac{\phi}{v}\right)\right]
\ee
where $f(0) = f'(0) = 0$. Restricting to positive values of $\phi$, inflation occurs as the inflaton
grows away from the origin. We denote by $\phi = v$ the minimum of the potential where $V=0$, $f(1) = 1$
and $f'(1) = 0$. The higher order derivatives of $f(x)$ at $x=1$ are usually of order one. In small field models,
inflation occurs when $\phi \ll v$~\footnote{If $v \gtrsim \mp$, inflation may end when the inflaton is relatively
close to the minimum of the potential $\phi \sim v$. This occurs for instance in modular inflation~\cite{modular} and
natural inflation~\cite{natural}. These models belong also to the hilltop category but they are not small-field models
(the inflaton takes instead Planckian values) and we do not consider them here. Preheating in these models is more
similar to preheating after chaotic inflation~\cite{KLS}.}. The inflationary potential may then be expanded into a
power series around the origin as
\be
\label{potor}
f\left(\frac{\phi}{v}\right) \, = \, \frac{\lambda}{p}\,\left(\frac{\phi}{v}\right)^p + ... \;\;\;\; \mbox{ for }
\;\; \phi \rightarrow 0
\ee
with $p > 1$ and $\lambda > 0$. The absence of the $p = 1$ term in (\ref{potor}) may follow naturally from the fact that
$\phi = 0$ could be a fixed point of some internal symmetries, but this is not always the case. The location of any fixed point
of internal symmetries is not important for the inflationary dynamics but it has crucial consequences for reheating, as we will see in Section~\ref{SecPert}. The original new inflation model \cite{new} is well described by the potential
(\ref{pot}, \ref{potor}) with $p=4$ (up to logarithmic corrections) and $\phi = 0$ a fixed point of internal symmetries. In the original scenario, the inflaton was supposed to be a Higgs field in Grand Unified Theories with $v \sim 10^{16}$ GeV, but this is not necessary. Another example is provided by the model of MSSM inflation of \cite{MSSM}, where the inflaton corresponds to a flat direction of the Minimal Supersymmetric Standard Model. This model is well described by the potential (\ref{pot}, \ref{potor}) with $p=3$ and $\phi=v$ a fixed point of internal symmetries~\footnote{In~\cite{MSSM}, the inflaton is normalized to $\phi = 0$ at the minimum of the potential.}, where $v \sim \left(m \, \Lambda^3\right)^{1/4}$ is related to the cutoff scale $\Lambda$ and to the soft SUSY breaking scale $m \sim$ TeV.

The details of preheating will of course depend on the explicit expression of the function $f(\phi/v)$ and in particular
on the form of the potential around its minimum. However, in this section, we will first review the inflationary dynamics, in order to determine the initial conditions for preheating and the constraint to impose on $M$ and $v$ from the normalization of the CMB fluctuations. For this, it is sufficient to consider the form (\ref{pot}, \ref{potor}) of the potential in the vicinity of $\phi = 0$. Let us first suppose that a single power of the inflaton $\phi^p$ in (\ref{potor}) dominates the potential throughout inflation. The slow-roll parameters then read
\bea
\label{epsilon}
\epsilon &\equiv& \frac{\mp^2}{2}\,\left(\frac{V'}{V}\right)^2 \simeq
\frac{\lambda^2}{2}\,\frac{\mp^2}{v^2}\,\left(\frac{\phi}{v}\right)^{2(p-1)} \\
\label{eta}
\eta &\equiv& \mp^2\,\left(\frac{V''}{V}\right) \simeq
-\lambda\,(p-1)\,\frac{\mp^2}{v^2}\,\left(\frac{\phi}{v}\right)^{p-2}
\eea
where a prime denotes derivative with respect to $\phi$ and $\mp = 1/\sqrt{8\pi G}$ is the reduced Planck mass.
It follows that $\epsilon / |\eta| \, \sim \, |\eta|\,\phi^2 / \mp^2$, so that $\epsilon \ll |\eta|$ during slow-roll
($|\eta| < 1$) for small field models ($\phi \ll \mp$). The slow-roll phase then ends when $|\eta| \sim 1$, while inflation continues afterwards until $\epsilon \sim 1$. Since $|\eta|$ must increase with $\phi$ for the slow-roll phase to end,
Eq.~(\ref{eta}) implies that the inflaton potential must be dominated by a power $\phi^p$ with $p > 2$ at the end of
slow-roll. This does not forbid the presence of terms with smaller powers of the inflaton in the potential
(\ref{pot},\ref{potor}) during inflation, but these terms must be subdominant at the end of slow-roll. In particular, a mass term $\phi^2$ is usually present in the potential when the model is embedded into supergravity and its coefficient must be somewhat smaller than its generic value in order to have $|\eta| < 1$ during inflation, the so-called
$\eta$-problem~\footnote{The $\eta$-problem in the context of small-field inflation is discussed in
e.g.~\cite{lowscale, pseudonatural}.}. It is possible that a mass term $\phi^2$ dominates the potential when cosmological scales leave the Hubble radius while a higher order term dominates at the end of inflation. However, it turns out that the constraint on $M$ and $v$ that follows from the normalization of CMB fluctuations in this case is practically the same as the one obtained when a single power $\phi^p$ with $p > 2$ dominates the potential throughout inflation, Eq.~(\ref{CMB}) below. Therefore, it is sufficient to consider a single power of the inflaton $\phi^p$ with $p > 2$ in the potential
(\ref{pot},\ref{potor}) in order to determine both the initial conditions for preheating and the constraint to impose on
$M$ and $v$ in order to obtain an acceptable spectrum of density perturbations.

Preheating will start at the end of the slow-roll phase, i.e. when $\eta \simeq -1$ but $\epsilon \ll 1$, because the potential term then starts to dominate in the equation of motion of the inflaton fluctuations, see the following Section.
Eq.~(\ref{eta}) gives
\be
\label{phiend}
\left(\frac{\phi_e}{v}\right)^{p-2} \simeq \frac{1}{\lambda (p-1)}\,\frac{v^2}{\mp^2}
\ee
for the value $\phi_e$ of the inflaton at the end of slow-roll when $\eta \simeq -1$. Using $3 H \dot{\phi} \simeq - V'$ leads to
\be
\label{dotphiend}
\dot{\phi}_e \simeq \frac{M^2}{\sqrt{3}\,\left[\lambda\,(p-1)^{p-1}\right]^{1/(p-2)}}\,\left(\frac{v}{\mp}\right)^{p/(p-2)}
\ee
for the time derivative of the inflaton at that time. These expressions provide the initial conditions
for the inflaton condensate at the beginning of preheating. However, inflation will continue afterwards because
$\epsilon \ll 1$ when $\eta \simeq -1$. Inflation ends when $\epsilon \simeq 1$, so that Eq.~(\ref{epsilon}) gives
\be
\label{phiepsilon}
\left(\frac{\phi_{\epsilon}}{v}\right)^{p-1} \simeq \frac{\sqrt{2}}{\lambda}\,\frac{v}{\mp}
\ee
for the value $\phi_{\epsilon}$ of the inflaton at the end of inflation. Note that $\phi_e \ll \phi_{\epsilon}$ for
$v \ll \mp$. When the inflaton condensate rolls from $\phi = \phi_e$ to $\phi = \phi_{\epsilon}$, preheating occurs while the universe is still inflating.

The value $\phi_N$ of the inflaton field $N$ e-folds before the end of inflation is given by
\be
\label{phiN}
\left(\frac{\phi_N}{v}\right)^{2-p} = \lambda\,(p-2)\,N\,\frac{\mp^2}{v^2}
\ee
which can be derived from $N = \mp^{-1}\,\int_{\phi_N}^{\phi_{\epsilon}} d\phi / \sqrt{2 \epsilon}$ where the integral is dominated by its lower limit. The normalization of the CMB fluctuations requires
\be
\label{CMB}
\left(\frac{V}{\epsilon}\right)_*^{1/4} \simeq 6.6 \, 10^{16}\,\mathrm{GeV} \equiv M_{\mathrm{infl}}
\ee
when the cosmological scales leave the Hubble radius at $N = N_*$. Using Eqs.~(\ref{epsilon},\ref{phiN}), this gives
\be
\label{CMBMv}
\frac{M^2}{v^2}\,\left(\frac{v}{\mp}\right)^{(p-4)/(p-2)} \simeq
\frac{\left[(p-2) N_*\right]^{-(p-1)/(p-2)}}{\sqrt{2}\,\lambda^{1/(p-2)}}\,\frac{M_{\mathrm{infl}}^2}{\mp^2} \ .
\ee
For instance, for $p = 3$ and $4$, this implies that $M \ll v$. Depending on the values of $v$ and $p$, inflation
can occur in a very wide range of energy scales. Note also that Eqs.~(\ref{CMB}) and (\ref{CMBMv}) hold
in the usual scenario where the curvature perturbation is generated from the vacuum fluctuations of the inflaton field.
If another field participates in the production the observed CMB fluctuations, like for
instance in the curvaton scenario~\cite{curvaton}, then the "$\simeq$" should be replaced by a "$<$" in these two equations. Finally, defining
\be
\label{defC}
C \equiv \sqrt{2}\,\lambda^{1/(p-2)}\,\left[(p-2) N_*\right]^{(p-1)/(p-2)}
\ee
the CMB normalization (\ref{CMBMv}) can be re-written
\be
\label{CMBv}
\frac{v}{\mp} = C^{(p-2)/p}\,\left(\frac{M}{M_{\mathrm{infl}}}\right)^{2(p-2)/p} \ .
\ee
For $\lambda \gtrsim 1$, Eq.~(\ref{defC}) gives $C > 1$. Eq.~(\ref{CMBv}) then implies that $M < M_{\mathrm{infl}}$ for any $p > 2$ and $v < \mp$.

If the curvature perturbation is generated by the quantum fluctuations of the inflaton field only and if the term
(\ref{potor}) with $p > 2$ dominates when cosmological scales leave the Hubble radius, the spectral index
$n_s \simeq 1 + 2 \eta - 6 \epsilon$ is given by
\be
\label{ns}
1 - n_s \simeq \frac{2 (p-1)}{(p-2) N_*} \ .
\ee
In small field models, inflation may occur at low energy scales and thermalization after preheating may be delayed (see Section~\ref{SecPert}), so that the number of e-folds $N_*$ before the end of inflation when cosmological scales leave the Hubble radius may be relatively small. This in turn may be constrained by the observational lower bound on $1-n_s$ via
Eq.~(\ref{ns}). However, contrary to Eq.~(\ref{CMBMv}), Eq.~(\ref{ns}) is significantly modified if a mass term in
$\phi^2$ dominates the potential when cosmological scales leave the Hubble radius, and the spectral index cannot be used to constrain the energy scale of inflation in that case.

Examples of small field inflation models are described by the effective potential
\be
\label{potp}
V = M^4\,\left(1 - \frac{\phi^p}{v^p}\right)^2
\ee
which reduces to (\ref{pot},\ref{potor}) with $\lambda = 2p$ during inflation. In \cite{lowscale}, models of this
kind with
\be
\label{vM}
v^p \approx M^q\,\mp^{p-q}
\ee
were constructed in supergravity, where $p > 2$ and $q$ are integers that depend on the symmetries of the theory.
Using (\ref{vM}) in (\ref{CMBMv}) shows that, even in the usual scenario for the origin of the cosmological
perturbations, inflation may occur at a very low energy scale $M$. The examples considered
in \cite{lowscale} include a case where $M$ corresponds to the intermediate scale of SUSY breaking in gravity-mediated scenarios, $M \sim 10^{11}$ GeV with $v \sim 10^{15}$ GeV, and a case with $M \sim v \sim 1$ GeV.

Let us now discuss the evolution of the inflaton condensate $\phi(t)$ starting from the end of slow-roll. It satisfies
the Klein-Gordon equation
\be
\label{KG}
\ddot{\phi} + 3 H \dot{\phi} + V'(\phi) = 0
\ee
where a dot denotes the derivative with respect to the cosmological time $t$, a prime denotes the derivative with
respect to $\phi$ and $H = \dot{a}/a$ is the Hubble rate. A quantity of prime interest is the curvature of the
potential $V''(\phi)$, which governs both the time scale for the variation of the inflaton condensate and the
typical scales that will be amplified by preheating. At the beginning of preheating $\phi = \phi_e$, $\eta \simeq -1$
so that
\be
\label{d2Ve}
V''(\phi_e) \simeq -3\,H_e^2 \simeq -\,\frac{M^4}{\mp^2}
\ee
where $H_e$ is the Hubble rate at the end of slow-roll. At the minimum of the potential, $\phi = v$, we have instead
\be
\label{d2Vm}
V''(v) \approx \frac{M^4}{v^2}
\ee
where we have taken $|f''(1)| \approx 1$. There must be an inflection point where $V'' = 0$ in between $\phi_e$
and $v$, and we will denote it by $\phi = \phi_i$. Since $|V''(\phi)|$ is increasing at the end of slow-roll, there is also a value of $\phi$ in between $\phi_e$ and $\phi_i$ where $|V''(\phi)|$ is maximum, and we will denote it as
$\phi = \phi_m$. Thus
\be
\label{phimi}
V'''(\phi_m) = 0 \;\;\;\;\; \mbox{ and } \;\;\;\;\; V''(\phi_i) = 0
\ee
with $\phi_e < \phi_m < \phi_i < v$. For $f''(1) \sim f'''(1) \sim 1$, we have $\phi_m \sim \phi_i \sim v$, whereas
$\phi_e \ll v$ for $v \ll \mp$. Note also that $|V''(\phi_e)| / V''(v) \approx v^2 / \mp^2 \ll 1$, so the inflaton
stays a long time close to $\phi_e$ before plunging rapidly towards the minimum of the potential.

The condensate then oscillates around the minimum with an amplitude that decreases with time due to Hubble friction.
The decrease can be estimated as in \cite{PreNew}, from the conservation of energy that follows from Eq.~(\ref{KG})
\be
\label{Econs}
\frac{1}{2}\,\dot{\phi}^2 + V(\phi) \, = \, \frac{1}{2}\,\dot{\phi}_i^2 + V(\phi_i) - \Delta V
\ee
where the dissipation term
\be
\label{dissip}
\Delta V = 3\,\int_{t_i}^t dt\,H\,\dot{\phi}^2
\ee
results from the Hubble friction. Denoting by $\phi_j$ the value of $\phi$ after $j$ complete oscillations of the inflaton (i.e. when $\phi_j$ is minimum and $\dot{\phi}_j = 0$, while $\phi_{j = 0} = \phi_e$), we have
\be
\label{Adec}
V(\phi_{j-1}) - V(\phi_j) = 3\,\int_{\phi_{j-1}}^{\phi_j} d\phi\,H\,\dot{\phi} \ .
\ee
The integral above is dominated by its contribution around $\phi \sim v$ and can be approximated by taking
$H \simeq M^2 / (\sqrt{3} \mp)$, $\dot{\phi} \approx M^2$ and $\Delta \phi \approx {2} v$. During the first oscillations of the inflaton, the potential at $\phi_j \ll v$ can be approximated by (\ref{pot},\ref{potor}).
Eq.~(\ref{Adec}) then gives
\be
\label{phin}
\frac{\phi_j}{v} \approx \left(j\,\frac{2\sqrt{3}\,p}{\lambda}\,\frac{v}{\mp}\right)^{1/p}
\ee
after $j \geq 1$ inflaton oscillations.

%%%%%%%%%%%%%%%%%%%%%%%%%%%%%%%%%%%%%%%%%%%%%%%%%%%%%%%%%%%%%%%%%%%%%%%%%%%%%%%%%

\section{First Stage of Preheating: Inflaton's Rolling}
\label{Sec1}

In this section, we  study the amplification of the inflaton fluctuations during the first stage of preheating,
when the inflaton condensate first rolls towards the minimum of the potential at the end of slow-roll, i.e. when
$\phi_e \leq \phi \leq v$. We then estimate the conditions under which preheating ends during this first stage. The
second stage of preheating (if preheating does not end during the first stage), when the inflaton condensate passes
the minimum of the potential and oscillates around it, occurs in a qualitatively different way. This second stage of preheating, and the subsequent decay of the inflaton fluctuations into different forms of particles, will be considered
in the following sections.

The first stage of preheating depends essentially on the form of the potential at the end of inflation, so we can consider the generic form (\ref{pot}, \ref{potor}) for small field inflation without specifying the form of the potential around its minimum. The typical scale of the fluctuations amplified during this first stage will be of the order
of the Hubble scale. In this case, the coupling of the inflaton fluctuations $\delta \phi$ to metric perturbations may
be non-negligible. During the linear stage of preheating, we can consider the Mukhanov variable
\be
\label{mukhanov}
\psi = \delta \phi + \frac{\dot{\phi}}{H}\,\Phi
\ee
where $\Phi$ is the metric perturbation in the longitudinal gauge
\be
ds^2 = -\left(1 + 2 \Phi\right)\,dt^2 + a^2(t)\,\left(1 - 2 \Psi\right)\,\vec{dx}^2
\ee
with $\Phi = \Psi$ (since the anisotropic stress vanishes at linear order in the perturbations). In Fourier space,
it satisfies the equation
\be
\label{eompsi}
\ddot{\psi}_k + 3 H\,\dot{\psi}_k + \left(\frac{k^2}{a^2} + V''(\phi) + 2\,\frac{\ddot{H}}{H}
- 2\,\frac{\dot{H}^2}{H^2} + 6\,\dot{H}\right)\,\psi_k = 0
\ee
where $k$ denotes the comoving wave-number. During inflation, we have the usual amplification of perturbations by the
quasi-exponential expansion of the scale factor
\be
\label{psite}
\psi_k \simeq \frac{H}{\sqrt{2\,k^3}}\,\left(i + \frac{k}{a\,H}\right)\,e^{ik / (aH)} \ .
\ee
This provides the initial conditions for the evolution of the perturbation during preheating.

Defining as usual $v = a^{3/2}\,\psi$, Eq.~(\ref{eompsi}) reduces to the equation of an harmonic oscillator
\be
\label{oscillator}
\ddot{v}_k + \omega_k^2(t)\,v_k = 0
\ee
with frequency squared
\be
\label{frequency}
\omega_k^2(t) = \frac{k^2}{a^2} + V''(\phi) + \Delta
\ee
where $\Delta \sim H^2$. Preheating starts when the potential term $V''(\phi)$ starts to dominate over $\Delta$ in this equation, i.e. at the end of the slow-roll phase when $\phi = \phi_e$, $\eta \simeq -1$ and $V'' \simeq -3 H^2$. Since
$V'' < 0$, the modes with $k^2 / a^2 < |V''|$ have a negative frequency squared and they are amplified by a tachyonic instability. During all the time that the condensate rolls from $\phi = \phi_e$ to $\phi = \phi_\epsilon \gg \phi_e$, see Eq.~(\ref{phiepsilon}), preheating occurs while the first slow-roll parameter $\epsilon$ is still much smaller than one and the universe is still inflating. As a consequence, $H$ is still approximately constant throughout the first stage of preheating. Eq.~(\ref{eompsi}) then reduces to the equation of motion for the inflaton fluctuations and the metric perturbation is negligible, $\psi \approx \delta \phi$. Furthermore, since the universe is still inflating, some of the modes that are amplified by the tachyonic instability leave the Hubble radius at the beginning of preheating. This is specific to small field inflation, where preheating starts when $V'' \sim -H^2$ and $\epsilon \ll 1$.

As the inflaton condensate rolls towards the minimum of the potential, all the modes with $k^2/a^2$ smaller than
$|V''(\phi)|$ are amplified by the tachyonic effect. This varies from $|V''(\phi_e)| \sim H^2$ at the end of inflation,
up to $|V''(\phi_m)|$ (the maximum of $|V''(\phi)|$, see (\ref{phimi})) which is typically of the order of
$|V''(\phi_m)| \sim V''(v) \sim M^4 / v^2 \gg H^2$, see (\ref{d2Vm}). For $v \ll \mp$, this represents a very wide
range of momenta. However, the process is dominated by the low-momentum modes, which are amplified during a much longer period of time and at a much higher rate. Indeed, the maximal growth occurs for the modes with
$k^2/a^2 \leq |V''| \simeq 3 H^2$ at the end of slow-roll, which are amplified from the beginning of preheating. In the following, we denote by $H_e \simeq M^2 / (\sqrt{3}\,\mp)$ the Hubble rate at the end of slow-roll / beginning of preheating($t = t_e$) and we normalize the scale factor as $a(t_e) = 1$. In this section, we will focus on the modes with
$k \lesssim H_e$, which dominates the first stage of preheating. The evolution of the modes with higher momenta, which we will see dominate the second stage of preheating, will be studied in the next section.

We can study the evolution of the low-momentum modes as in \cite{PreNew}, by noticing that they behave as the velocity of
the homogeneous inflaton condensate $\dot{\phi}$, which is given by the conservation of energy (\ref{Econs}). Indeed,
the time derivative of Eq.~(\ref{KG}) gives
\be
\label{eomy}
\ddot{y} + 3H\,\dot{y} + \left(V''(\phi) + 3\,\dot{H}\right)\,y = 0
\ee
where $y = \dot{\phi}$. This reduces to Eq.~(\ref{eompsi}) for $k^2 \lesssim H_e^2 \ll |V''|$ (except at the very
beginning of preheating and very close to the inflection point where $V''$ vanishes),
$\ddot{\psi}_k + V''\,\psi_k \simeq 0$. The general solution for $\psi_k$ with $k \lesssim H_e$ then reads
\be
\psi_{k \lesssim H_e} \simeq c_1\,\dot{\phi} + c_2\,\dot{\phi}\,\int^t \frac{dt}{\dot{\phi}^2}
\ee
where $c_1$ and $c_2$ are constants. As the inflaton rolls from $\phi = \phi_e$ to $\phi = v$, $\dot{\phi} \neq 0$ and the integral above is a slowly varying function of time, so that we have essentially $\psi_k \propto \dot{\phi}$. Note that this implies in particular that, for $H$ constant, the comoving curvature perturbation
$\mathcal{R} = H\,\psi / \dot{\phi} \propto H$ is conserved for super-Hubble modes, as it should. Indeed, for $H$ constant and $k/a \rightarrow 0$, Eqs.~(\ref{eompsi}) and (\ref{eomy}) are identical and $\psi \propto \dot{\phi}$ exactly.

\begin{figure}[htb]
\begin{center}
\includegraphics[width=10cm]{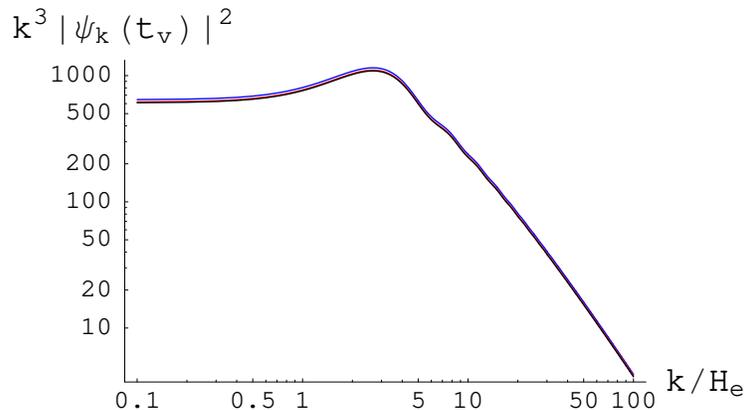}
\end{center}
\vspace*{-5mm}
\caption{Spectra $\left(\frac{v}{\mp}\right)^4\,H_e^{-2}\,k^3\,|\psi(t_v)|^2$ as a function of $k/H_e$ in the linear regime of preheating for the model (\ref{potp}) with $p = 4$ and $v = 10^{16}$ GeV, $10^{13}$ GeV and $10^{10}$ GeV. The $3$ spectra lie on top of each other and are in very good aggreement with Eq.~(\ref{k3psi2}).}
\label{k3psi2tv}
\vspace*{-3mm}
\end{figure}

The initial velocity of inflaton condensate at the beginning of preheating, $t = t_e$, is given in (\ref{dotphiend}).
Let us denote by $t = t_v$ the  time when the inflaton condensate passes for the first time by the minimum
of the potential $\phi = v$. Eq.~(\ref{Econs}) gives $\dot{\phi}(t_v) \simeq \sqrt{2}\,M^2$. Therefore
\be
\label{psitv}
\frac{\psi_k(t_v)}{\psi_k(t_e)} \simeq \frac{\dot{\phi}(t_v)}{\dot{\phi}(t_e)} \simeq
\sqrt{6}\,\left[\lambda\,(p-1)^{p-1}\right]^{1/(p-2)}\,\left(\frac{\mp}{v}\right)^{p/(p-2)}
\ee
for the low-momentum modes. For $v \ll \mp$, the growth of the perturbation from $t = t_e$ to $t = t_v$ is very large.
Using Eq.(\ref{psite}) for $\psi_k(t_e)$, Eq.~(\ref{psitv}) gives
\be
\label{k3psi2}
k^3\,|\psi_k(t_v)|^2 \simeq 3\,\left[\lambda\,(p-1)^{p-1}\right]^{2/(p-2)}\,\left(\frac{\mp}{v}\right)^{2 p/(p-2)}\,
H_e^2\,\left(1 + \frac{k^2}{H_e^2}\right)
\ee
for $k \lesssim H_e$. We have solved the equations of motion (\ref{KG}, \ref{eompsi}) for the inflaton condensate and the perturbations numerically for the potential (\ref{potp}) with $p = 4$, starting from the initial conditions at the end of slow-roll. The spectrum of $k^3\,|\psi_k(t_v)|^2$ for different values of $v$ is shown in Fig.~\ref{k3psi2tv}. For the
low-momentum modes $k \lesssim H_e$, which dominate the spectrum, the results agree very well with the analytical prediction (\ref{k3psi2}). The amplitude of the spectrum is maximum at $k_* \simeq 4\,H_e$ and then quickly decreases
for larger values of $k$ (this part of the spectrum will be studied in the next section). Because $a H$ is still growing
at the beginning of preheating, the peak of the spectrum at the end of the first stage is in fact slightly outside the Hubble radius, $k_* / (a\,H) \simeq 0.3$.

Eq.~(\ref{k3psi2}) is valid in the linear regime of preheating, when the amplitude of the inflaton fluctuations is much smaller than the inflaton condensate, $|\delta \phi(t_v)| \ll v$. In that case, the energy density at $t = t_v$ is still dominated by the inflaton condensate, which starts oscillating around the minimum of the potential. Otherwise, the backreaction of the inflaton fluctuations shuts off the tachyonic effect and preheating ends in less than one oscillation of the condensate. We can estimate the variance of the inflaton fluctuations
$\delta \phi \approx \psi$ at $t = t_v$ from Eq.~(\ref{k3psi2}),
$\langle \delta \phi^2 \rangle \propto \int d\mathrm{ln}k \,k^3\,|\psi_k|^2$. As usual, the integral must be regularized
or restricted to a given range of scales due to the mild logarithmic divergence of the inflationary perturbations in the IR (in addition of course to the usual power law divergence of the vacuum fluctuations in the far UV). The integral is then dominated by the peak of the spectrum at $k \sim H_e$. This gives
\be
\langle \delta \phi^2(t_v) \rangle \approx \left[k^3\,|\psi_k(t_v)|^2\right]_{k \sim H_e} \approx
\left[\lambda\,(p-1)^{p-1}\right]^{2/(p-2)}\,H_e^2\,\left(\frac{\mp}{v}\right)^{2 p/(p-2)}
\ee
where we have neglected factors of order unity. Preheating does not end during the first stage if
$\langle \delta \phi^2(t_v) \rangle < v^2$, i.e.
\be
\label{cond2}
\left(\frac{v}{\mp}\right)^{1/(p-2)}\,>\,\left[\lambda\,(p-1)^{p-1}\right]^{1/(2p-4)}\,\frac{M}{v} \ .
\ee
In that case, the subsequent oscillations of the inflaton condensate lead to a second stage of preheating, as discussed in the next section.	

On the other hand, if the condition (\ref{cond2}) is not satisfied, preheating ends already during the first stage. In that case, the spectrum of the inflaton fluctuations at the end of preheating is peaked around the Hubble scale, or even slightly outside the Hubble radius, before the energy gets redistributed by rescattering. As mentioned in the introduction, large field fluctuations at the Hubble scale are rather specific to preheating after small field inflation and may have interesting consequences for the production of primordial black holes and gravitational waves. Let us now discuss the kind of models where this occurs, i.e. for which the condition (\ref{cond2}) is violated and preheating ends in less than one oscillation of the condensate. In the model of New Inflation \cite{new}, $p = 4$, $M/v \sim 10^{-3}$ and $\lambda \sim 1$, so preheating ends in less than one oscillation for $v / \mp \lesssim 10^{-6}$ \cite{PreNew}. In the model of MSSM inflation of
\cite{MSSM}, $p = 3$, $\lambda \sim 1$, $M \sim \sqrt{m\,v}$ and $v \sim \left(m\,\Lambda^3\right)^{1/4}$ where $m$ is a soft SUSY breaking mass and $\Lambda$ the cutoff scale (denoted as $\Lambda^{n-3} = M_P^{n-3} / \lambda_n$ in \cite{MSSM}). In that case, the condition (\ref{cond2}) is violated, and therefore the spectrum of inflaton fluctuations is peaked around the Hubble scale at the end of preheating, for
$\Lambda / \mp \lesssim \left(m / \mp\right)^{1/9}$. For $m \sim$ TeV, this occurs for
$\Lambda / \mp \lesssim 10^{-2}$, in particular for $\Lambda$ of the order of the GUT scale.

In general, we can also determine the condition for preheating to end during the first stage in terms of $v$ or $M$ only by using their relation (\ref{CMBMv}) from the CMB normalization. Eq.~(\ref{cond2}) is then violated for
\be
\label{cond1v}
\frac{v}{\mp} \lesssim \left[\frac{p-1}{(p-2) N_*}\right]^{(p-1)/(p-2)}\,\left(\frac{M_{\mathrm{infl}}}{\mp}\right)^2 \ .
\ee
Using (\ref{CMBv}), this upper bound on $v$ gives an upper bound on $M$, so that preheating ends during the first stage if inflation occurs at a sufficiently low energy scale. This upper bound on $M$ is very sensitive to the value of $p$, see Section \ref{SecPert} for details. In terms of $v$, the condition (\ref{cond1v}) for preheating to end in less than one oscillation of the inflaton condensate is roughly $v/\mp \lesssim 10^{-6} - 10^{-5}$.

%%%%%%%%%%%%%%%%%%%%%%%%%%%%%%%%%%%%%%%%%%%%%%%%%%%%%%%%%%%%%%%%%%%%%%%%%%%%%%%%%

\section{Second Stage of Preheating: Inflaton Oscillations}
\label{Sec2}

When the condition (\ref{cond2}) is satisfied, the inflaton condensate still carries most of the energy density at the end of the first stage of preheating and it starts to oscillate around the minimum of the potential. In this section, we study the evolution of the inflaton fluctuations during this second stage of preheating when $v / \mp > 10^{-6}$. The decay of the inflaton into other fields will be studied in the next section.

In order to study the second stage of preheating quantitatively, we need to specify the form of the potential around its minimum. We will first consider the potential (\ref{potp}) with $p=4$
\be
\label{potp4}
V = M^4\,\left(1 - \frac{\phi^4}{v^4}\right)^2 \ .
\ee
The generalization to other models of small field inflation will be discussed in sub-section~\ref{SubSecGen}.
For the model (\ref{potp4}), the value $\phi_e$ of the condensate at the end of slow-roll, (\ref{phiend}),
the inflection point $\phi_i$ and the point $\phi_m$ where $- V''$ is maximum, see (\ref{phimi}), are given by
\be
\label{phimiex}
\phi_e \simeq \frac{v^2}{\sqrt{24}\,\mp} \;\;\;\;\;\; ,  \;\;\;\;\;\; \phi_m = \left(\frac{1}{7}\right)^{1/4}\,v
\;\;\;\;\;\; \mbox{ and }  \;\;\;\;\;\; \phi_i = \left(\frac{3}{7}\right)^{1/4}\,v \, .
\ee
Note again that $\phi_e \ll \phi_m \lesssim \phi_i \lesssim v$. The mass of the inflaton at the minimum of the potential reads
\be
\label{mmin}
m_{\mathrm{min}} = \sqrt{V''(v)} = 4 \sqrt{2}\,\frac{M^2}{v}
\ee
and the maximum momentum of the inflaton fluctuations amplified by the tachyonic effect is
\be
\label{kmax}
\frac{k_{\mathrm{max}}}{a} = \sqrt{-V''(\phi_m)} = \frac{4}{7^{1/4}}\,\frac{M^2}{v} \simeq 0.435\,m_{\mathrm{min}} \ .
\ee

To study the evolution of the inflaton fluctuations $\delta \phi \approx \psi$, it is convenient to consider their
comoving occupation number
\be
\label{nk}
n_k = \frac{\omega_k}{2}\,\left(|v_k|^2 + \frac{|\dot{v}_k|^2}{\omega_k^2}\right) - \frac{1}{2}
\ee
which is conserved when the frequency $\omega_k(t)$ varies adiabatically with time (and is real). Here the modes $v_k$ satisfy the oscillator equation (\ref{oscillator}) with the frequency (\ref{frequency}). Although the occupation number is only well-defined when $\omega_k^2 > 0$, it is convenient to generalize the definition (\ref{nk}) to the case where
$\omega_k^2 < 0$ too, by using an "effective frequency squared" given by (\ref{frequency}) where $V''$ is replaced by its absolute value. This reduces to the correct expression for the occupation number when $\omega_k^2 > 0$ and it allows to also follow the evolution of the modes when they are tachyonic.

\begin{figure}[htb]
\begin{tabular}{cc}
\begin{minipage}[t]{8cm}
\begin{center}
\includegraphics[width=8cm]{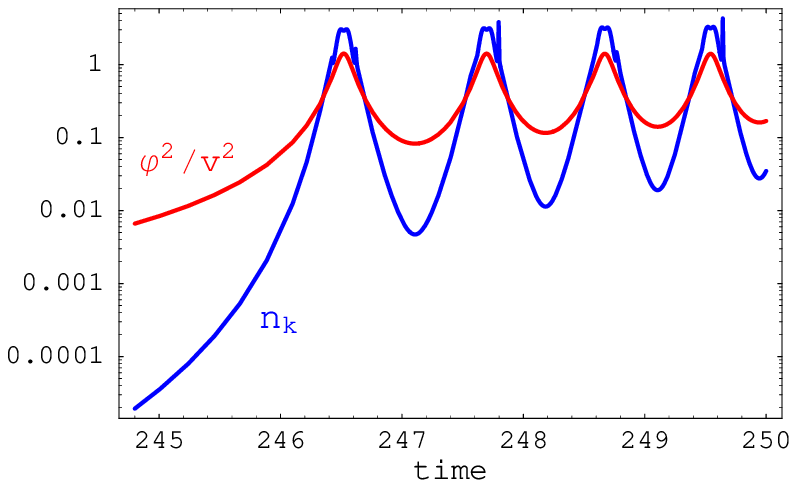}
\caption{Evolution with time of the occupation number $n_k$ for the mode with $k = H_e$ (blue) and of the square of the inflaton condensate $\phi^2(t) / v^2$ (red), for the model (\ref{potp4}) with expansion of the universe and $v = 10^{16}$ GeV. Here $n_k$ has been rescaled by an arbitrary factor in order to fit on the plot. Note that the plot does not display the evolution from the end of inflation but focus instead on the first oscillations of the condensate.}
\label{phink}
\end{center}
\end{minipage}&
\hspace*{1.5cm}
\begin{minipage}[t]{8cm}
\begin{center}
\includegraphics[width=8cm]{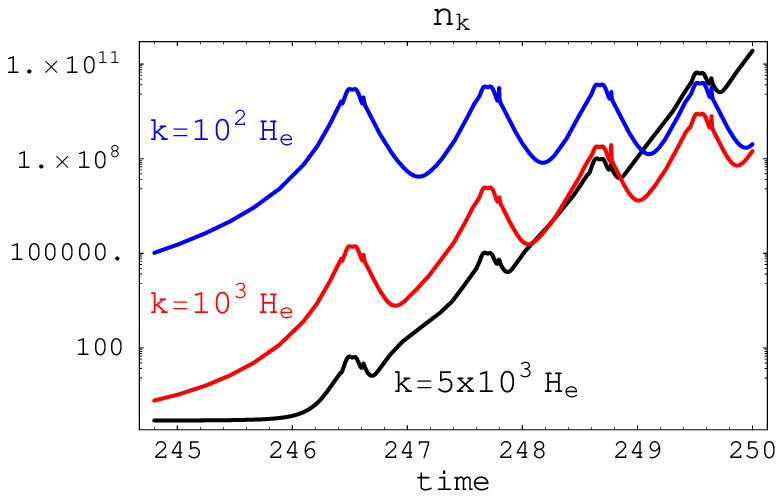}
\caption{Evolution with time of the occupation number $n_k$ for the modes with $k = 10^2\,H_e$ (blue), $k = 10^3\,H_e$ (red) and $k = 5 \times 10^3\,H_e$ (black), for the model (\ref{potp4}) with expansion of the universe and $v = 10^{16}$ GeV.}
\label{nk3k}
\end{center}
\end{minipage}
\end{tabular}
\end{figure}

We solved numerically Eq.~(\ref{oscillator}) for the model (\ref{potp4}) during the linear stage of preheating for different values of $k$ and $v / \mp$, and we computed the occupation number as defined above. In Fig.~\ref{phink}, we show the evolution with time of $n_k$ for $k = H_e$ and of the square of the inflaton condensate $\phi^2(t) / v^2$, during the first oscillations of the condensate, for $v = 10^{16}$ GeV. The condensate first rolls slowly from $\phi = \phi_e \ll v$ and then oscillates rapidly around the minimum $\phi = v$. During each oscillation, $\phi$ varies between a maximum value
$\phi \simeq 2^{1/4}\,v$ and a minimum value $\phi_j$ which agrees well with (\ref{phin}) and which increases at each oscillation due to Hubble friction. The main observation from Fig.~\ref{phink} is that, for the modes with $k \sim H_e$ which dominate the first stage of preheating, while the occupation number $n_k$ grows exponentially by a large amount when the condensate first rolls towards $\phi = v$, it also {\it decreases} exponentially by a large amount when the condensate rolls back towards the origin. In other words, the amplification of this mode during the first tachyonic episode is partly canceled by the second tachyonic episode. It would be fully canceled if during the second tachyonic episode the inflaton condensate rolled back up to the value from where it started. The process then repeats itself as the condensate rolls back towards the minimum and oscillates. Overall, the occupation number of the modes with $k/a \lesssim H$ oscillates with a large amplitude during the second stage of preheating, with almost no net growth after the first stage. Indeed, these low-momentum modes still follow the behaviour of the condensate velocity,
$v_k \propto \dot{\phi}$ and $\dot{v}_k \propto \ddot{\phi} \propto V'(\phi)$. When $\phi$ is maximum, the first term in (\ref{nk}) vanishes but the second one is maximum, so that $n_k$ is maximum. When $\phi$ is minimum, the first term vanishes and the second one is small, so that $n_k$ is minimum.

In Fig.~\ref{phink}, we show the evolution with time of the occupation number for modes with higher momenta. All these modes have $k < k_{\mathrm{max}}$ in the tachyonic regime and are amplified by the tachyonic instability. As we discussed in the previous section, the modes with higher momenta grow less when the condensate first rolls towards the minimum of the potential. However, we see in Fig.~\ref{phink} that these modes also decrease less when the condensate goes back towards the origin, so they progressively catch up with the modes with smaller momenta. Overall, there is a competition between
low-momentum modes which grow a lot but also decrease a lot and modes with higher momenta which grow less but also decrease less~\footnote{A similar process has been observed numerically for a potential $V = V_0 - m^2\,\phi^2 + \lambda\,\phi^4$ in the context of preheating after hybrid inflation in the second paper of \cite{tachyonic}.}. As the condensate oscillates around the minimum of the potential, the spectrum of the inflaton fluctuations is peaked at much larger momenta than during the first stage of preheating and the peak is progressively shifted towards the UV. Such a UV peak was attributed in
\cite{PreNew} to a non-adiabatic production of the inflaton fluctuations, but we will see that it follows rather from the large exponential decrease of the low-momentum modes when their frequency squared is negative.

We will call {\it tachyonic oscillation} this succession of exponential increases and decreases of the low-momentum modes with $k < k_{\mathrm{max}}$. In order to study this process in more detail, let us first neglect the expansion of the universe, $a(t) = 1$. In that case, the inflaton condensate starts from some initial value $\phi_0 \ll v$, with an initial velocity that we can take to vanish, $\dot{\phi}_0 = 0$. It then evolves periodically in time and comes back to its initial value $\phi_0$ after each oscillation. It is convenient to rescale the field, the time coordinate and the wave-numbers as
\be
\label{tildes}
\tilde{\phi} = \frac{\phi}{v} \;\;\;\;\;\; ,  \;\;\;\;\;\; \tilde{t} = \frac{m_{\mathrm{min}}}{2}\,t
\;\;\;\;\;\; ,  \;\;\;\;\;\; \tilde{k} = \frac{2}{m_{\mathrm{min}}}\,k \ .
\ee
The conservation of energy (\ref{Econs}) with $\dot{\phi}_0 = 0$ then reads
\be
\label{Etilde}
\frac{1}{2}\,\dot{\tilde{\phi}}^2 + \tilde{V}(\tilde{\phi}) \, = \, \tilde{V}(\tilde{\phi_0})
\ee
where a dot now denotes the derivative with respect to the {\it rescaled} time $\tilde{t}$ and we have defined the
rescaled potential
\be
\label{Vtilde}
\tilde{V} = \frac{1}{8}\,\left(1 - \tilde{\phi}^4\right)^2 \ .
\ee
In terms of the rescaled variables (\ref{tildes}), the mode equation for the inflaton fluctuations (\ref{oscillator})
without expansion of the universe reads
\be
\label{modeq}
\ddot{v}_k + \tilde{\omega}_k^2(\tilde{t})\,v_k = 0 \;\;\;\;\;\; \mbox{ with }  \;\;\;\;\;\;
\tilde{\omega}_k^2(\tilde{t}) = \tilde{k}^2 + \tilde{V}''\left(\tilde{\phi}(\tilde{t})\right)
\ee
where a prime now denotes the derivative with respect to $\tilde{\phi}$. Defining
\be
\label{ketilde}
\tilde{k}_0 = \sqrt{-\tilde{V}''(\phi_0)} \simeq \sqrt{3}\,\tilde{\phi}_0 \, ,
\ee
the modes with $\tilde{k} < \tilde{k}_0$ have a negative frequency squared initially and are amplified by the tachyonic instability from the beginning. On the other hand, the maximum wave-number amplified by the tachyonic instability
(\ref{kmax}) is
\be
\label{kmaxtilde}
\tilde{k}_{\mathrm{max}} = \left(\frac{4}{7}\right)^{1/4} \simeq 0.87 \ .
\ee
In the rest of this section, we will always work with the rescaled variables (\ref{tildes}, \ref{Vtilde}, \ref{modeq}),
but from now on we omit the tilde on $\tilde{\phi}$, $\tilde{t}$, $\tilde{k}$ and $\tilde{\omega}_k$ to simplify notations.

\begin{figure}[htb]
\begin{center}
\includegraphics[width=13cm]{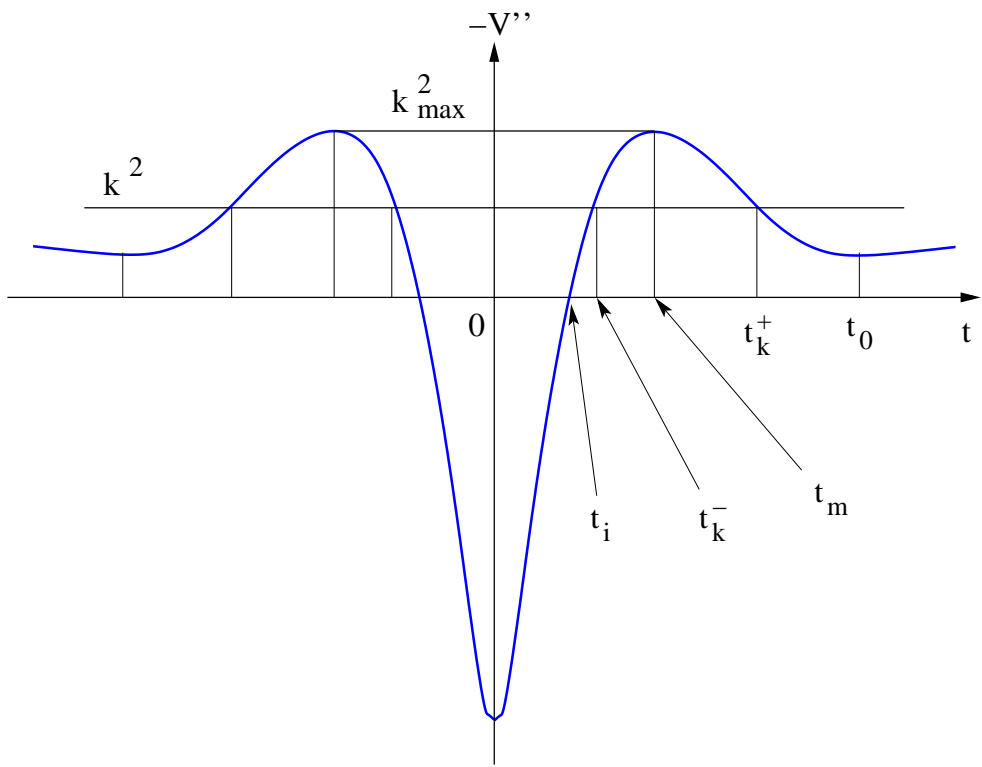}
\end{center}
\vspace*{-5mm}
\caption{The shape of the effective Schr\"{o}dinger potential $-\tilde{V}''\left(\phi(t)\right)$ in Eq.~(\ref{modeq}) for
the model (\ref{Vtilde}) during one oscillation of the inflaton condensate. Here for convenience the coordinate $t$ has been normalized to $t = 0$ at the center of the "crater" - i.e. when the inflaton condensate reaches its maximum value before oscillating back. The full potential is made of a periodic succession of these volcano potentials during each oscillation of the inflaton condensate.}
\label{volcano}
\vspace*{-3mm}
\end{figure}

The mode equation (\ref{modeq}) can be viewed as a Schr\"{o}dinger equation in one spatial dimension $t$, where the
time evolution of the modes $v_k(t)$ corresponds to the spatial profile of the wave-functions of states with energy $k^2$
in the effective potential $-\tilde{V}''\left(\phi(t)\right)$. For the model (\ref{Vtilde}), the shape of this effective Schr\"{o}dinger potential during one oscillation of the inflaton condensate is shown in Fig.~\ref{volcano}. The full potential is made of a periodic succession of these potentials during each oscillation of the inflaton condensate.
We see from Fig.~\ref{volcano} that the Schr\"{o}dinger potential has a volcano shape, with two "crests" separated by a deep "crater". The time evolution of the modes $v_k(t)$ during one oscillation of the inflaton condensate corresponds to a tunneling through this potential for $k < k_{\mathrm{max}}$ and to a scattering above this potential for
$k \geq k_{\mathrm{max}}$. In this language, the large exponential decrease of the low-momentum modes that we observed above when the inflaton condensate rolls back towards small values of $\phi$ corresponds to a resonant tunneling of low-energy states~\footnote{Note in this respect that a decrease of the modes in time can be viewed as an increase of the wave-functions in space, as the orientation of the coordinate $t$ corresponding to the time-evolution of the modes is "opposite" to the one corresponding to the spatial profile of the wave-functions. Indeed, for the time evolution of the modes, the initial conditions correspond to only positive-frequency waves ($\alpha_k = 1$ and
$\beta_k = 0$ in (\ref{alpha})) in the vacuum, while in general both positive- and negative-frequency waves are excited after one condensate oscillation when particles are produced. In the opposite way, in the quantum mechanical problem of tunneling through a potential barrier, both incident and reflected waves are present before the barrier, while only transmitted waves are present after the barrier.}. Indeed, the wave-function of these low-energy states is localized at the center of the deep "crater" and decreases symmetrically away from it.

Let us first consider the time evolution of the modes $v_k(t)$ during a single oscillation of the inflaton condensate, starting from $\phi = \phi_0 \ll 1$. We will normalize the time coordinate such that $t = 0$ corresponds to the  time after half a period of oscillation, when the inflaton condensate reaches its maximum value before oscillating back towards small values of $\phi$. We denote by $t = -t_0$ the time when the inflaton condensate starts to roll away from its initial value $\phi = \phi_0$, by $t = -t_m$ the moment when it reaches the value $\phi = \phi_m$ and by
$t = -t_i$ the moment when it reaches the value $\phi = \phi_i$, see (\ref{phimiex}). Thus we have $0 < t_i < t_m < t_0$, see Fig.~\ref{volcano}. Since here we neglect the expansion of the universe, the time evolution is symmetric around
$t = 0$: as the inflaton rolls back towards small values of $\phi$ after $t = 0$, it takes the same values $\phi = \phi_i$, $\phi_m$ and $\phi_0$ at the times $t = t_i$, $t_m$ and $t_0$ respectively. For the modes with $k \geq k_{\mathrm{max}}$, the frequency squared $\omega_k^2(t)$ in (\ref{modeq}) is always positive. For the modes with $k < k_{\mathrm{max}}$, it is negative during a $k$-dependent interval of time in the tachyonic regime. We will denote this time interval by
$\left[-t_k^+ , -t_k^-\right]$ when $t < 0$ and $\left[t_k^- , t_k^+\right]$ when $t > 0$, see Fig.~\ref{volcano}. Thus
\be
\label{phikpm}
k^2 + \tilde{V}''(\phi_k^\pm) = 0
\ee
where $\phi_k^{\pm} = \phi(\pm t_k^\pm)$. The modes with $k < k_0$, Eq.~(\ref{ketilde}), have a negative frequency squared from the begining of preheating, so that $t_k^+ = t_0$ for these modes.

A convenient way of solving the mode equation, that is often useful in the context of preheating (see
e.g.~\cite{KLS, tachres}), is to use the WKB approximation when the frequency varies adiabatically with time, i.e. when
$|\dot{\omega}_k| \ll \omega_k^2$ and $|\ddot{\omega}_k| \ll \omega_k^3$ (including imaginary values of $\omega_k$ when
$\omega_k^2 < 0$). Here however, we find that these conditions are satisfied only in the intervals of time
$\left[-t_0 , -t_m\right]$ and $\left[t_m , t_0\right]$. Let us first consider the evolution of the modes in these intervals of time. For the modes with $k_0 < k < k_{{\mathrm{max}}}$, see Eqs.~(\ref{ketilde},\ref{kmaxtilde}), we have
$\omega_k^2(t) > 0$ for $-t_0 \leq t < -t_k^+$ and for $t_k^+ < t \leq t_0$. The solution of Eq.~(\ref{modeq}) can then be written as a superposition of positive and negative frequency waves
\be
\label{alpha}
v_k(t) \simeq \frac{\alpha_k}{\sqrt{2\,\omega_k(t)}}\,\mathrm{exp}\left(-i \int_{-t_0}^t dt'\, \omega_k(t')\right)
+ \frac{\beta_k}{\sqrt{2\,\omega_k(t)}}\,\mathrm{exp}\left(i \int_{-t_0}^t dt'\, \omega_k(t')\right)
\ee
for $-t_0 \leq t < -t_k^+$ and
\be
\label{alphabar}
v_k(t) \simeq \frac{\bar{\alpha}_k}{\sqrt{2\,\omega_k(t)}}\,\mathrm{exp}\left(-i \int_{t_k^+}^t dt'\, \omega_k(t')\right)
+ \frac{\bar{\beta}_k}{\sqrt{2\,\omega_k(t)}}\,\mathrm{exp}\left(i \int_{t_k^+}^t dt'\, \omega_k(t')\right)
\ee
for $t_k^+ < t \leq t_0$. The Bogolyubov coefficients $\alpha_k$, $\beta_k$, $\bar{\alpha}_k$ and $\bar{\beta}_k$ are constant when the frequency evolves adiabatically with time. For the modes with $k < k_{{\mathrm{max}}}$, when the frequency squared is negative the solution of (\ref{modeq}) can be written as a superposition of exponentially increasing and decreasing solutions
\be
\label{a}
v_k(t) \simeq \frac{a_k}{\sqrt{2\,\Omega_k(t)}}\,\mathrm{exp}\left(-\int_{-t_k^+}^t dt'\, \Omega_k(t')\right)
+ \frac{b_k}{\sqrt{2\,\Omega_k(t)}}\,\mathrm{exp}\left(\int_{-t_k^+}^t dt'\, \Omega_k(t')\right)
\ee
for $-t_k^+ < t \leq -t_m$ and
\be
\label{abar}
v_k(t) \simeq \frac{\bar{a}_k}{\sqrt{2\,\Omega_k(t)}}\,\mathrm{exp}\left(-\int_{t_m}^t dt'\, \Omega_k(t')\right)
+ \frac{\bar{b}_k}{\sqrt{2\,\Omega_k(t)}}\,\mathrm{exp}\left(\int_{t_m}^t dt'\, \Omega_k(t')\right)
\ee
for $t_m \leq t < t_k^+$, where $\Omega_k^2 = - \omega_k^2$ is positive. Again, the coefficients $a_k$, $b_k$,
$\bar{a}_k$ and $\bar{b}_k$ are constant in the adiabatic regime. The second term in Eqs.~(\ref{a},\ref{abar}) is  exponentially growing with time and corresponds to the tachyonic amplification of the modes. However, if the coefficient of the first, exponentially decreasing term is much larger than the coefficient of the exponentially increasing one, then the mode will start to decrease exponentially with time and this will explain the behaviour observed in Figs.~\ref{phink} and \ref{nk3k}.

The WKB approximation is not valid in the vicinity of the turning points $t = \pm t_k^+$, where the frequency vanishes, but there the oscillatory and exponentially evolving solutions can be matched, as described below (see also \cite{tachres}), and this allows to follow the evolution of the modes through these moments of time too. However, the WKB approximation is not applicable around the entire crater of the Schr\"{o}dinger potential, i.e. when $-t_m < t < t_m$, because the potential varies more abruptly there. Thus, in that region, it does not seem possible to find an approximate solution for the exact
Schr\"{o}dinger potential. In order to follow the evolution of the modes in that region, our strategy will be instead to find the exact solution for an approximate Schr\"{o}dinger potential. To sum up, we will solve the mode equation
(\ref{modeq}) for the modes with $k < k_{{\mathrm{max}}}$ as follows: (i) use the WKB approximations
(\ref{alpha} - \ref{abar}) in the intervals of time $\left[-t_0 , -t_m\right]$ and $\left[t_m , t_0\right]$, (ii) match the solutions (\ref{alpha}) with (\ref{a}) and (\ref{alphabar}) with (\ref{abar}) around the turning points $t = \pm t_k^+$ for the modes with $k > k_0$, (iii) find an approximate Schr\"{o}dinger potential in the interval $\left[-t_m , t_m\right]$ for which the Schr\"{o}dinger equation can be solved analytically and (iv) match this exact solution with the WKB approximations (\ref{a}, \ref{abar}) at $t = \pm t_m$.

We can apply the same procedure for the modes with $k \geq k_{\mathrm{max}}$, except that in this case we will use the oscillatory solutions (\ref{alpha}, \ref{alphabar}) in the WKB approximation in the entire intervals
$\left[-t_0 , -t_m\right]$ and $\left[t_m , t_0\right]$, since the frequency squared is always positive for these modes. These oscillatory solutions will then be matched to the exact solution of the approximate Schr\"{o}dinger equation in the interval $\left[-t_m , t_m\right]$. Note that, although the modes with $k \geq k_{\mathrm{max}}$ are not amplified by the tachyonic instability, they can still be amplified by the non-adiabatic evolution of the frequency in the interval
$\left[-t_m , t_m\right]$. We will see however that this effect is in fact negligible.

The method outlined above allows one to follow the evolution of the modes and requires that we can indeed use the WKB approximation at
$t = \pm t_m$, i.e. that the frequency varies adiabatically at these moments in time. This is very well satisified for the modes with $k \ll k_{\mathrm{max}}$, which dominate the second stage of preheating, because
$\dot{\Omega}_k \propto V'''(\phi)\,\dot{\phi}$ vanishes at $t = \pm t_m$ since $V'''(\phi_m) = 0$. Similarly, the WKB approximation is very good at $t = \pm t_m$ for the modes with $k \gg k_{\mathrm{max}}$. However, it is not applicable for the modes with $k \approx k_{\mathrm{max}}$, because the frequency vanishes around $t = \pm t_m$ for these modes. In principle, we could describe the evolution of these modes by finding another approximation of the Schr\"{o}dinger potential that is accurate also outside the interval $\left[-t_m , t_m\right]$. However, this is not necessary, since we will see that our results are accurate even for $k$ only slightly less or slightly bigger than $k_\mathrm{max}$, and the behaviour of the modes with $k \approx k_{\mathrm{max}}$ can be easily understood as an interpolation between these two regimes.

\begin{figure}[htb]
\begin{center}
\includegraphics[width=9cm]{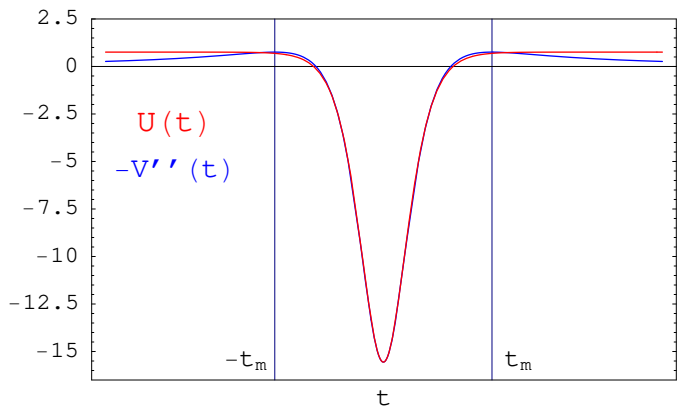}
\end{center}
\vspace*{-5mm}
\caption{Comparison between the exact function $-\tilde{V}''(t)$ (in blue) calculated numerically and the analytical approximation $U(t)$ (in red) defined in (\ref{Upot}). The agreement is very good in the interval
$\left[-t_m , t_m\right]$.}
\label{Uapp}
\vspace*{-3mm}
\end{figure}

It remains to find a Schr\"{o}dinger potential which is a good approximation of $-\tilde{V}''\left(\phi(t)\right)$ in the interval $\left[-t_m , t_m\right]$ and for which exact analytical solutions of the Schr\"{o}dinger equation can be found. As discussed in the Appendix, a good choice is given by
\be
\label{Upot}
-\tilde{V}''\left(\phi(t)\right) \simeq U(t) = k_{{\mathrm{max}}}^2 - \frac{U_0}{\cosh^2(\mu t)}
\ee
with
\be
\label{mumag}
\mu = \sqrt{\frac{U_0}{2} + \frac{k_{{\mathrm{max}}}^2}{16}} - \frac{3}{4}\,k_{{\mathrm{max}}} \ .
\ee
The above relation between the parameters does not follow from the only requirement that $U(t)$ gives a good
approximation to $-\tilde{V}''$. Indeed, we show in the Appendix that $\mu$ should satisfy (\ref{mumag}) in order to reproduce the known evolution of the zero mode $v_{k = 0} \propto \dot{\phi}$. We also show there that for the model
(\ref{potp}) with $p = 2$, the {\it exact} Schr\"{o}dinger potential $-\tilde{V}''$ appearing in the mode equation is given by (\ref{Upot}, \ref{mumag}) with other numerical values of the parameters. As discussed in Section \ref{SecBack}, this model does not lead to small field inflation, but in that case both the background evolution of the inflaton condensate and the wave-equation for the perturbations can be exactly solved analytically. For the model (\ref{potp4}), $\phi$ is maximum with $\dot{\phi} = 0$ at $t = 0$, so Eq.~(\ref{Etilde}) gives $\phi = 2^{1/4}$ and therefore $-\tilde{V}'' = 11 \sqrt{2}$. This is reproduced by (\ref{Upot}) for
\be
U_0 = k_{{\mathrm{max}}}^2 + 11 \sqrt{2} \simeq 16.3
\ee
where we have used (\ref{kmaxtilde}). Because $\mu\,t_m >> 1$, $U(t)$ reproduces also the correct value
$k_{{\mathrm{max}}}^2$ of $-\tilde{V}''$ at $\phi = \phi_m$. The potential $U(t)$ with the above values of the parameters is shown in Fig.~\ref{Uapp}. We see that it agrees very well with the numerical solution for
$-\tilde{V}''\left(\phi(t)\right)$ in the interval $\left[-t_m , t_m\right]$.

In the Appendix, we solve the Schr\"{o}dinger equation for the potential $U(t)$ and match the solution to the WKB approximations (\ref{alpha}-\ref{abar}). In the next sub-section, we use these results to study the second stage of preheating for the modes with $k < k_{{\mathrm{max}}}$. In sub-section~\ref{SubSecG}, we do the same for the modes with
$k > k_{{\mathrm{max}}}$. We then study the effect of the expansion of the universe on the second stage of preheating in sub-section~\ref{SubSecExp}. Finally, we generalize our results to other models of small field inflation in
sub-section~\ref{SubSecGen}.

\subsection{"Tachyonic Oscillations" for $k < k_{{\mathrm{max}}}$}
\label{SubSecS}

Consider first the modes with $k < k_{{\mathrm{max}}}$, still without the expansion of the universe. In this case,
we have to match the exact solution of the approximate Schr\"{o}dinger equation in the interval $\left[-t_m , t_m\right]$ to the WKB approximations (\ref{a}) and (\ref{abar}). As shown in the Appendix, this leads to the transfer matrix
\be
\label{abaram}
\left(\begin{array}{c}
\bar{a}_k \vspace*{0.1cm}\\
\bar{b}_k
\end{array}\right) = e^{X_k}\,
\left(\begin{array}{cc}
0 & - 1 \vspace*{0.1cm}\\
e^{-2 X_k} & E\,k^2 /k_\mathrm{max}^2
\end{array}\right) \,
\left(\begin{array}{c}
a_k \vspace*{0.1cm}\\
b_k
\end{array}\right)
\ee
between the solutions (\ref{a}) and (\ref{abar}) in the limit $k \ll k_{{\mathrm{max}}}$. We will see however that this approximation is valid even when $k$ is only slightly smaller than $k_{{\mathrm{max}}}$. Here $E \simeq 2.15$ is a numerical constant and
\be
\label{Xk}
X_k = \int_{t_m}^{t_k^+} \Omega_k(t)\,dt =
\int_{\phi_k^+}^{\phi_m} \frac{\sqrt{|\tilde{V}''(\phi)| - k^2}}{|\dot{\phi}|}\,d\phi
\ee
is the "exponential phase" accumulated during half a period of the condensate oscillation when $\omega_k^2 < 0$ and the WKB approximation is applicable. The lower limit of the integral above depends on the wave-number.
The modes with $k \leq k_0 \simeq \sqrt{3}\,\phi_0$ - see Eq.~(\ref{ketilde}) - have a negative frequency squared from the beginning, so that $\phi_k^+ = \phi_0$ for these modes. On the other hand, for the modes with $k > k_0$, $\phi_k^+$ is the turning point where the frequency vanishes. For $k \ll k_\mathrm{max}$, we have $\phi_k^+ \ll 1$ and Eqs.~(\ref{Vtilde}, \ref{phikpm}) then give $\phi_k^+ \simeq k /\sqrt{3}$. Thus
\be
\label{phik+}
\phi_k^+ \simeq \left\{\begin{array}{ccc}
\frac{k}{\sqrt{3}} & \mbox{ for } & k > \sqrt{3}\,\phi_0 \vspace*{0.1cm} \\
\phi_0 & \mbox{ for } & k \leq \sqrt{3}\,\phi_0
\end{array} \right.
\ee
for the modes with $k \ll k_\mathrm{max}$.

It is interesting to understand the behaviour of different modes when the condensate climbs back the potential from
$\phi \sim 1$ towards $\phi = \phi_0$. For the modes with $k / k_{{\mathrm{max}}} \lesssim e^{-X_k}$, which includes in particular the zero-mode $k = 0$, Eq.~(\ref{abaram}) gives $\bar{a}_k \sim - e^{X_k}\,b_k$ and
$\bar{b}_k \sim e^{- X_k}\,a_k$. Plugging into (\ref{abar}), we then have $v_k(t_k^+) \sim (b_k - a_k) / \sqrt{2 \Omega_k}$ which, from (\ref{a}), is of the order of $v_k(- t_k^+)$. Thus we recover the fact that the low-momentum modes oscillate without any net growth during one oscillation of the condensate, because $v_k \propto \dot{\phi}$ for these modes. Indeed,
for the modes with $k / k_{{\mathrm{max}}} \lesssim e^{-X_k}$, we have $|\bar{a}_k / \bar{b}_k| \sim e^{2 X_k}$ and the solution (\ref{abar}) decreases exponentially during the whole interval of time $\left[t_m , t_k^+\right]$. Now we consider the modes with $e^{-X_k} \ll k / k_{{\mathrm{max}}} \ll 1$, for which the factor in $e^{-2 X_k}$ in (\ref{abaram}) is neglibile. For these modes, (\ref{abaram}) gives $|\bar{a}_k / \bar{b}_k| \sim k^2 / k_\mathrm{max}^2$, which is still very large so that the solution (\ref{abar}) will again start to decrease exponentially with time, but not  enough for this solution to decrease all the way until $t = t_k^+$.

In order to follow the evolution of the modes with $k > k_0$ during a full oscillation of the inflaton condensate, it remains to match the eponential solutions (\ref{a},\ref{abar}) with the oscillatory solutions (\ref{alpha},\ref{alphabar}) around the turning points $t = \pm t_k^+$, where the frequency vanishes and the WKB approximation breaks down. If we formally extend the domain of definition of the functions $v_k(t)$, $\omega_k(t)$ and $\Omega_k(t)$ to the complex plane
$t$, then the matching can be done by going around the real turning points along a contour in the complex plane which is sufficiently far away from the turning points that the WKB approximation remains valid along the contour~\cite{LL}. This procedure leads to the transfer matrix (see ~\cite{tachres} for more details)
\be
\label{aalpha}
\left(\begin{array}{c}
a_k \vspace*{0.1cm}\\
b_k
\end{array}\right) = e^{-i \pi /4}\,
\left(\begin{array}{cc}
i e^{-i \Theta_k} &  e^{i \Theta_k} \vspace*{0.1cm}\\
e^{-i \Theta_k} & i e^{i \Theta_k}
\end{array}\right) \,
\left(\begin{array}{c}
\alpha_k \vspace*{0.1cm}\\
\beta_k
\end{array}\right)
\ee
between the solutions (\ref{alpha}) and (\ref{a}) around $t = -t_m$, where
\be
\label{Thetak}
\Theta_k = \int_{t_k^+}^{t_0} \omega_k(t)\,dt =
\int_{\phi_0}^{\phi_k^+} \frac{\sqrt{k^2 + \tilde{V}''(\phi)}}{|\dot{\phi}|}\,d\phi
\ee
is the "oscillatory phase" accumulated during half a period of the condensate oscillation when $\omega_k^2 > 0$ and the
WKB approximation is applicable. Proceding similarly around $t = t_m$ gives
\be
\label{baralphaa}
\left(\begin{array}{c}
\bar{\alpha}_k \vspace*{0.1cm}\\
\bar{\beta}_k
\end{array}\right) = e^{-i \pi /4}\,
\left(\begin{array}{cc}
e^{- X_k} &  i e^{X_k} \vspace*{0.1cm}\\
i e^{- X_k} & e^{X_k}
\end{array}\right) \,
\left(\begin{array}{c}
\bar{a}_k \vspace*{0.1cm}\\
\bar{b}_k
\end{array}\right) \, .
\ee
The factors in $e^{-X_k}$ in this transfer matrix are again negligible for $k / k_{{\mathrm{max}}} > e^{-X_k}$ and
will be set to zero in the following.

Combining (\ref{abaram}), (\ref{aalpha}) and (\ref{baralphaa}), we get the full tranfer matrix between the solutions
(\ref{alpha}) and (\ref{alphabar}) during a complete oscillation of the inflaton condensate. We will denote by
$\alpha_k^{j}$ and $\beta_k^{j}$ the coefficients in (\ref{alpha}) {\it after} the $j^\mathrm{th}$ complete oscillation of the condensate (i.e. at the beginning of the $(j + 1)^\mathrm{th}$ oscillation) and by $\bar{\alpha}_k^{j-1}$ and
$\bar{\beta}_k^{j-1}$ the coefficients in (\ref{alphabar}) {\it during} the $j^\mathrm{th}$ oscillation of the condensate.
We then have the further relations $\alpha_k^{j} = \bar{\alpha}_k^{j-1}\,e^{-i \Theta_k}$ and
$\beta_k^{j} = \bar{\beta}_k^j\,e^{i \Theta_k}$. We can then relate the Bogolyubov coefficients after and before the
$j^\mathrm{th}$ oscillation of the inflaton condensate as
\be
\label{fulltransfer}
\left(\begin{array}{c}
\alpha_k^j \vspace*{0.1cm}\\
\beta_k^j
\end{array}\right) = E\,\frac{k^2}{k_\mathrm{max}^2}\,e^{2 X_k}\,
\left(\begin{array}{cc}
e^{- 2 i \Theta_k} &  i \vspace*{0.1cm}\\
- i & e^{2 i \Theta_k}
\end{array}\right) \,
\left(\begin{array}{c}
\alpha_k^{j-1} \vspace*{0.1cm}\\
\beta_k^{j-1}
\end{array}\right) \ .
\ee

Because here we neglect the expansion of the universe, all the coefficients in this transfer matrix are the same during each oscillation of the condensate. The occupation number of the inflaton fluctuations $n_k^j = |\beta_k^j|^2$ after $j$ complete oscillation of the inflaton condensate is then obtained by taking the $j^\mathrm{th}$ power of the tranfer matrix. With the initial conditions $\alpha_k^0 = 1$ and $\beta_k^0 = 0$ corresponding to vacuum fluctuations, we finally get
\be
\label{nkj}
n_k^j = \left[E \,\frac{k^2}{k_\mathrm{max}^2}\,e^{2 X_k}\right]^{2j}\,
\left[2\,\cos(2 \Theta_k)\right]^{2 (j-1)} \ .
\ee
It is interesting to compare this simple analytical expression to the one obtained in \cite{tachres} for the process of "tachyonic resonance" that takes place during preheating after chaotic inflation with trilinear interactions. In that case, the frequency squared of the modes becomes also periodically negative during some $k$-dependent intervals of time, but the modes are only growing exponentially during these tachyonic episodes. In Eq.~(\ref{nkj}), $2 X_k$ and $2 \Theta_k$ correspond to the total phase of $\Omega_k$ and $\omega_k$ accumulated in the adiabatic regime during a complete oscillation of the inflaton condensate, see (\ref{Xk}) and (\ref{Thetak}). The expression (\ref{nkj}) is then formally similar to the one obtained in \cite{tachres}, except for the crucial pre-factor in $k^2/k_\mathrm{max}^2$ that takes into account the effect of the temporary exponential decrease of the modes that occurs in preheating after small field inflation.

As in \cite{tachres}, after more than one oscillation of the condensate, the modes with $\cos(2 \Theta_k) = 0$ are not amplified because of destructive interferences between the successive tachyonic instabilities. These modes form narrow stability bands, separated by wide instability bands where the growth of the modes is governed by the factor in
$k^2\,e^{2 X_k}$ in Eq.~(\ref{nkj}). To calculate $X_k$, we can use Eqs.~(\ref{Etilde},\ref{Vtilde}) in the second
equality of (\ref{Xk}). For $k \ll k_\mathrm{max}$, the integral is dominated by its lower limit $\phi_k^+ \ll 1$.
This gives
\be
\label{EvalXk}
X_k \approx -\sqrt{6}\,\int^{\phi_k^+} \frac{\sqrt{\phi^2 - k^2 / 3}}{\sqrt{\phi^4 - \phi_0^4}}\,d\phi
\ee
where we neglected higher powers of $\phi \ll 1$ inside the integral. The value of $\phi_k^+$ is given in
Eq.~(\ref{phik+}). For $k \gg k_0$, we can neglect the term in $\phi_0^4$ in (\ref{EvalXk}). This gives
\be
\label{k2eXk}
\frac{k^2}{k_\mathrm{max}^2}\,e^{2 X_k} \approx \left(\frac{k}{k_\mathrm{max}}\right)^{-2 (\sqrt{6} - 1)}
\ee
for $k_0 \ll k \ll k_\mathrm{max}$. Thus, for these modes, the spectrum of the occupation number (\ref{nkj}) after $j$ oscillations of the condensate decreases with $k$ as $k$ to the power $-2 j (\sqrt{6} - 1)$. On the other hand, for
$k \ll k_0$, $\phi_k^+ = \phi_0$ and $X_k$ is approximately constant, so that $n_k^j$ grows as $k^{4j}$. In particular, the peak of the spectrum is located in between these two different regimes, at $k \simeq k_0$. Note also form (\ref{k2eXk}) that $k / k_\mathrm{max} \gg e^{-X_k}$ for $k \gtrsim k_0$. This implies that only the modes with $k \ll k_0$
(i.e. well below the peak of the spectrum) have non-negligible factors in $e^{-2X_k}$ in the transfer matrices
(\ref{abaram}) and (\ref{baralphaa}). Therefore, as discussed in the paragraph below (\ref{abaram}), only the modes with
$k \ll k_0$ decrease exponentially during the whole interval of time $\left[t_m , t_0\right]$. We will see that the expansion of the universe will modify this result.

\begin{figure}[htb]
\begin{center}
\includegraphics[width=5.9cm]{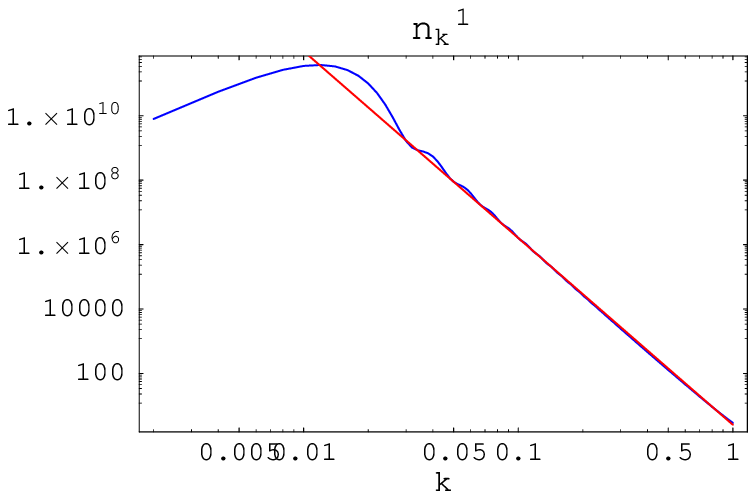}
\includegraphics[width=5.9cm]{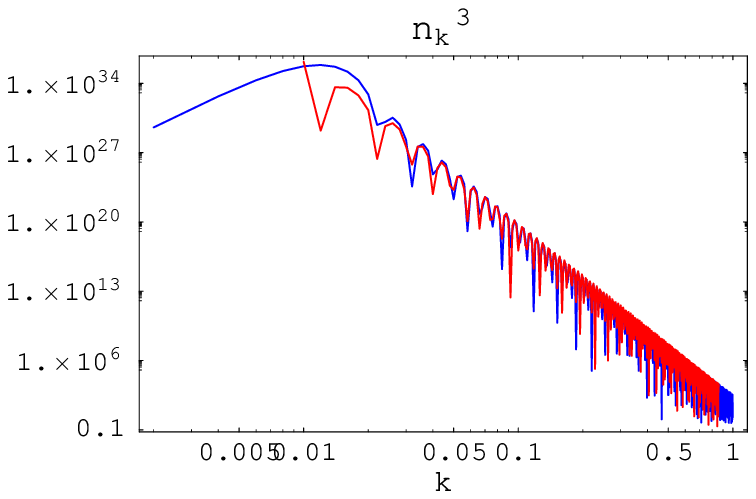}
\includegraphics[width=5.9cm]{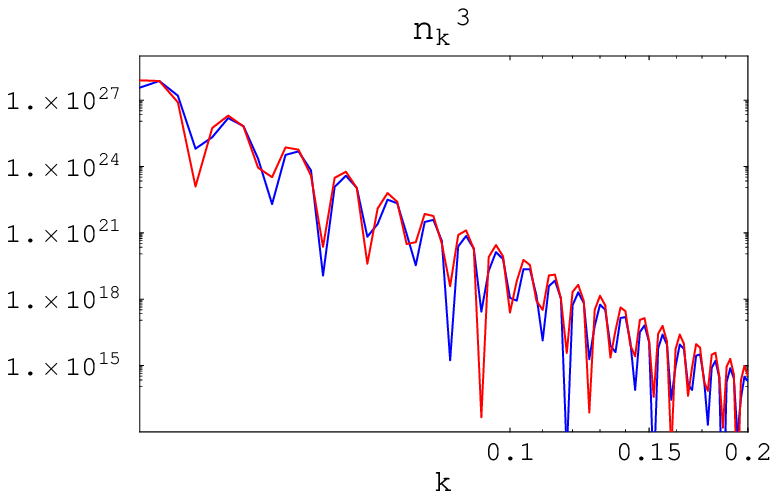}
\caption{Spectrum of the occupation number of the inflaton fluctuations (without expansion of the universe) after $j$ complete oscillations of the condensate, for $j = 1$ (left pannel) and $j = 3$ (middle and right pannel). Here we took
$\phi_0 = 10^{-2}$ so that $k_0 \simeq 0.017$, see (\ref{ketilde}), while $k_\mathrm{max} \simeq 0.87$, see
(\ref{kmaxtilde}). We compare the analytical expression (\ref{nkj}) (in red) with the spectrum obtained by solving the mode equation numerically (in blue). The right pannel is a zoom of the middle pannel on a shorter range of momenta.}
\label{nk13}
\end{center}
\end{figure}

The spectrum of the occupation number of the inflaton fluctuations after $j$ oscillations of the condensate is shown in Fig.~\ref{nk13}, where we compare the result obtained by solving numerically the mode equation to the analytical result (\ref{nkj}) in the range $k_0 \leq k$. The aggreement is very good from the peak at $k \simeq k_0$ up to
$k \sim k_\mathrm{max}$. The left pannel shows the spectrum after $j = 1$ oscillation of the condensate, while the middle and right pannels are for $j = 3$. In
the second case, we have narrow "stability bands" where $\cos(2 \Theta_k) = 0$ in (\ref{nkj}). When the expansion of the universe is taken into account, the physical momenta $k / a$ "move" inside the resonance pattern. This makes the band structure less distinct, but still present as we will see in sub-section~\ref{SubSecExp}.

\subsection{Non-Adiabatic Production for $k > k_{{\mathrm{max}}}$}
\label{SubSecG}

For the modes with $k > k_{{\mathrm{max}}}$, the frequency squared remains always positive but the inflaton fluctuations are still amplified in the interval of time $\left[-t_m , t_m\right]$ where the frequency varies non-adiabatically with time.
To study this case, we match the WKB solutions (\ref{alpha}) and (\ref{alphabar}) to the exact solution of the approximate Schr\"{o}dinger equation in the interval $\left[-t_m , t_m\right]$. As shown in the Appendix, we can then relate the occupation numbers of the inflaton fluctuations $n_k^j$ and $n_k^{j-1}$ after and before
the $j^\mathrm{th}$ oscillation of the condensate as
\be
\label{nkD}
n_k^j = |D_k|^2 + \left(1 + 2\,|D_k|^2\right)\,n_k^{j-1} +
2 |D_k|\,\sqrt{1 + |D_k|^2}\,\sqrt{n_k^{j-1}\,(n_k^{j-1} + 1)}\,\cos \theta_k^j
\ee
where $\theta_k^j$ is a phase defined in the Appendix and
\be
\label{Dk2}
|D_k| = \frac{\sin\left(\pi k_{{\mathrm{max}}} / \mu\right)}{\sinh\left(\pi \sqrt{k^2 - k_{\mathrm{max}}^2} / \mu\right)}
\, .
\ee
Eq.~(\ref{nkD}) is formally identical to the one obtained in \cite{KLS} for the non-adiabatic production of particles after chaotic inflation, but in that case $|D_k|^2 = e^{-\pi k^2 / k_*^2}$ where $k_*$ is the typical momentum amplified by the process.

\begin{figure}[htb]
\begin{tabular}{cc}
\begin{minipage}[t]{8cm}
\begin{center}
\includegraphics[width=8cm]{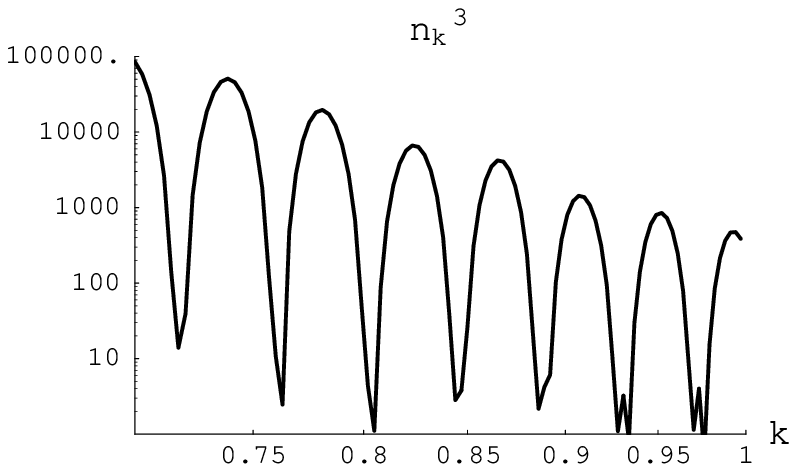}
\caption{Spectrum of the occupation number of the inflaton fluctuations (without expansion of the universe) after $3$ oscillations of the condensate, for $k \approx k_{\mathrm{max}} \simeq 0.87$.}
\label{nkmax}
\end{center}
\end{minipage}&
\hspace*{1.5cm}
\begin{minipage}[t]{8cm}
\begin{center}
\includegraphics[width=8cm]{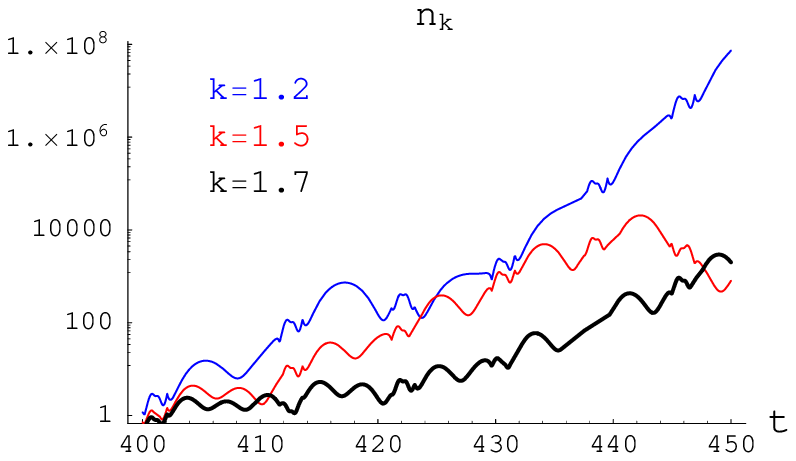}
\caption{Evolution with time of the (comoving) occupation number $n_k$ for modes with $k \approx k_{\mathrm{max}}$,
for the model (\ref{potp4}) with expansion of the universe and $v = 10^{16}$ GeV. The evolution is shown during the first
$6$ oscillations of the condensate, where the scale factor varies from $a \simeq 1.5$ to $a \simeq 1.7$
(with $a = 1$ at the beginning of preheating). Thus the maximum comoving momentum amplified by the tachyonic effect,
$k_{\mathrm{max}} \simeq 0.87\,a$, varies from $k_{\mathrm{max}} \simeq 1.3$ to $k_{\mathrm{max}} \simeq 1.5$. We show the evolution of the modes with comoving momenta (from top to bottom): $k = 1.2$ (blue), $k = 1.5$ (red) and $k = 1.7$ (black).}
\label{nkmaxexp}
\end{center}
\end{minipage}
\end{tabular}
\end{figure}

For $k \gg k_{\mathrm{max}}$, $|D_k| \propto e^{-\pi k / \mu}$ in (\ref{nkD}) and the non-adiabatic production is very inefficient. On the other hand, as $k$ decreases towards $k_{\mathrm{max}}$, $|D_k|$ in (\ref{Dk2}) can become smaller than one and the efficiency of the non-adiabatic production is enhanced. This is to be contrasted with the non-adiabatic production after chaotic inflation studied in \cite{KLS}, where $|D_k|^2 = e^{-\pi k^2 / k_*^2} < 1$. However, as discussed in the paragraph before Eq.~(\ref{Upot}), the matching method and therefore Eq.~(\ref{nkD}) are not valid for
$k \approx k_{\mathrm{max}}$. In this limit, there is a smooth transition between the tachyonic amplification for
$k < k_{\mathrm{max}}$ and the non-adiabatic production for $k > k_{\mathrm{max}}$. Indeed, when $k$ increases towards
$k_{\mathrm{max}}$, the frequency squared becomes negative in shorter and shorter intervals of time around $\pm t_m$ and the efficiency of the tachyonic amplification decreases considerably, whereas the efficiency of the non-adiabatic production increases when $k$ decreases towards $k_{\mathrm{max}}$. This is illustrated in Fig.~\ref{nkmax}, where we show the spectrum of the occupation number of the inflaton fluctuations for $k \approx k_{\mathrm{max}}$ after $3$ oscillations of the condensate. Note again the stability bands where destructive interferences occur. Their positions depend on
$\Theta_k$ in (\ref{nkj}) for $k < k_{\mathrm{max}}$ and on $\theta_k^j$ in (\ref{nkD}) for $k > k_{\mathrm{max}}$. Note also that, even if the non-adiabatic production is enhanced for $k \approx k_{\mathrm{max}}$, it is still much less efficient than the tachyonic amplification of the modes with $k \ll k_{\mathrm{max}}$ discussed in the previous
sub-section.

For the modes with $k \gtrsim k_{\mathrm{max}}$, particle production occurs on the physical (un-rescaled) time scale
$t_m \sim m_{\mathrm{min}}^{-1} \ll H^{-1}$, during which the expansion of the universe is negligible. For these modes,
the main effect of the expansion of the universe is that the physical momentum $k/a$ varies from one event of particle to the other. In that case, as in \cite{KLS}, the physical momenta move inside the resonance pattern of
Fig.~\ref{nkmax} during successive oscillations of the condensate. This implies in particular that the occupation number of a given comoving mode varies in a different way at different events of particle production. This is illustrated in
Fig.~\ref{nkmaxexp}, where we show the evolution with time of the (comoving) occupation number in an expanding universe for modes with $k \approx k_{\mathrm{max}}$. Because of the last term in (\ref{nkD}), the occupation numbers can decrease during one oscillation of the condensate, see e.g. the mode with $k = 1.5$ in Fig.~\ref{nkmaxexp}. Furthermore, since $k/a$ decreases with time whereas $k_{\mathrm{max}}/a = \sqrt{-V''(\phi_m)}$ remains constant, more and more modes enter the regime with $k / a \ll \sqrt{-V''(\phi_m)}$ where the tachyonic amplification becomes more and more efficient. This occurs for instance for the mode with $k = 1.2$ in Fig.~\ref{nkmaxexp}, which start with $k/a \approx \sqrt{-V''(\phi_m)}$ and ends with $k/a \ll \sqrt{-V''(\phi_m)}$. In the following sub-section, we discuss the effects of the expansion of the universe for the modes with $k \ll k_{\mathrm{max}}$, which dominate the second stage of preheating.

\subsection{Tachyonic Oscillations in an Expanding Universe}
\label{SubSecExp}

We now discuss how the process of "tachyonic oscillations" studied in sub-section~\ref{SubSecS} for the
modes with $k \ll k_{\mathrm{max}}$ is modified by the expansion of the universe. For these modes, the main effect of the expansion of the universe comes from the fact that the amplitude of the condensate oscillations decrease with time due to Hubble friction. The effect of Hubble friction on the evolution of the condensate was discussed in Section~\ref{SecBack}, where we obtained the estimate (\ref{Adec}) for the minimum value $\phi_j$ of the condensate when it climbs back the potential after $j \geq 1$ complete oscillation(s). For the model (\ref{potp4}), it reads
\be
\label{phij}
\tilde{\phi}_{j \geq 1} \approx \left(j\,\frac{v}{\mp}\right)^{1/4} \ .
\ee
Here we continue to use the rescaled variables (\ref{tildes}), but now we write the "tildes" explicitly to avoid confusion. In these variables, the initial value (\ref{phimiex}) of the condensate at the beginning of preheating reads
\be
\label{phij0}
\tilde{\phi}_{j = 0} = \tilde{\phi}_e \simeq \frac{v}{\sqrt{24}\,\mp} \ .
\ee
Thus $\tilde{\phi_j}$ is of the same order of magnitude for different values of $j \geq 1$, but
$\tilde{\phi}_{j \geq 1} \gg \tilde{\phi}_e$.

\begin{figure}[htb]
\begin{center}
\includegraphics[width=9cm]{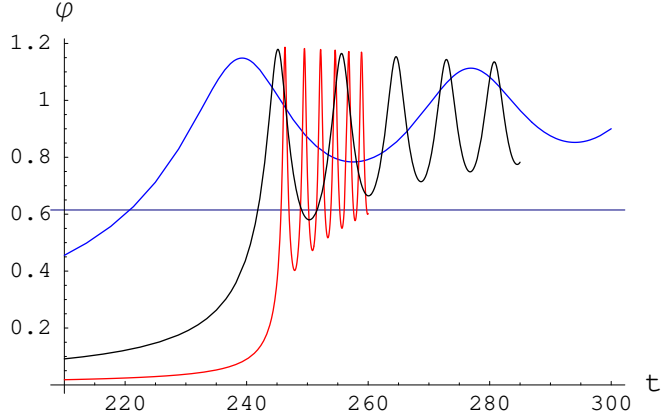}
\end{center}
\vspace*{-5mm}
\caption{Time evolution of the condensate $\tilde{\phi}(\tilde{t})$ for different values of $v / \mp$. From top to bottom at the left: $v / \mp = 0.5$ (blue), $0.1$ (black) and $0.02$ (red). The horizontal line shows
$\tilde{\phi} = \tilde{\phi_m} = 1/7^{1/4}$.}
\label{phimexp}
\vspace*{-3mm}
\end{figure}

First of all, as discussed in \cite{PreNew}, Hubble friction may be so important that the condensate never goes back in the tachyonic region $\phi < \phi_m$ after his first rolling down. Using Eqs.~(\ref{phimiex}) and (\ref{phij}),
$\tilde{\phi}_1 > \tilde{\phi}_m$ for
\be
\label{condnopre}
\frac{v}{\mp} \gtrsim 10^{-1} \ .
\ee
This is illustrated in Fig.~\ref{phimexp}, where we show the evolution with time of the inflaton condensate during its first oscillations for different values of $v / \mp$. When the condition (\ref{condnopre}) is satisfied, the process of tachyonic oscillation does not take place. The tachyonic amplification of inflaton fluctuations during the first stage of preheating is also rather inefficient for such high values of $v / \mp$. In that case, the main decay channel for the inflaton condensate is the decay into other fields, either perturbatively or non-perturbatively. The decay of the inflaton into other fields will be discussed in the next Section.

We also saw above that, for the model (\ref{potp4}), preheating ends after the first stage (i.e. in less than one oscillation of the condensate) for $v / \mp \lesssim 10^{-6}$. Thus we now discuss the effects of the expansion of the universe on the second stage of preheating for $10^{-6} < v / \mp < 10^{-1}$. In order to do so, it is convenient to define the analogue of (\ref{Xk}, \ref{phik+}) during the $j^{\mathrm{th}}$ oscillation of the condensate
\be
X_k^j = \int_{\tilde{\phi}_{k\,j}^{+}}^{\tilde{\phi}_m} \frac{\sqrt{|\tilde{V}''| - \tilde{k}^2/a^2}}{|\dot{\tilde{\phi}}|}\,d\tilde{\phi}
\ee
where
\be
\label{phikj+}
\tilde{\phi}_{k\,j}^+ \simeq \left\{ \begin{array}{ccc}
\frac{\tilde{k}}{\sqrt{3}\,a} & \mbox{ for } & \frac{\tilde{k}}{a} > \sqrt{3}\,\tilde{\phi}_j \vspace*{0.1cm} \\
\tilde{\phi}_j & \mbox{ for } & \frac{\tilde{k}}{a} \leq \sqrt{3}\,\tilde{\phi}_j \, .
\end{array} \right.
\ee
The scale factor $a$ does not vary much during the first few oscillations of the condensate after the first one. We will take it to be constant and equal, say to its value after the first complete oscillation of the condensate. Note also that we consider the modes with $\tilde{k} \ll \tilde{k}_{\mathrm{max}} \simeq 0.87$, so that $\tilde{k} / a \ll 1$. Proceding as in (\ref{EvalXk}, \ref{k2eXk}), we then have
\be
\label{eXkj}
e^{X_k^j} \sim \left\{\begin{array}{ccc}
\left(\frac{\tilde{k}}{a}\right)^{-\sqrt{6}} & \mbox{ for } & \frac{\tilde{k}}{a} > \sqrt{3}\,\tilde{\phi}_j
\vspace*{0.1cm} \\
\mbox{constant } \sim \tilde{\phi}_j^{-\sqrt{6}} & \mbox{ for } & \frac{\tilde{k}}{a} \leq \sqrt{3}\,\tilde{\phi}_j
\end{array} \right.
\ee
where we work at the level of an order of magnitude estimate. The main effect comes from the fact that $X_k^j$ now depends on time.

To see this, consider the $j^\mathrm{th}$ oscillation of the condensate, where $\phi$ starts from $\phi_{j-1}$, oscillates around the minimum of the potential and climbs back the potential up to $\phi_j$. We can then use the transfer matrix
(\ref{abaram}) with $X_k$ replaced by $X_k^{j-1}$, $\tilde{k}^2$ replaced by the physical momentum $\tilde{k}^2 / a^2$ and
$E / \tilde{k}_{\mathrm{max}}^2 \sim 1$. On the other hand, in the transfer matrix (\ref{baralphaa}), $X_k$ should be replaced by $X_k^j$. This leads, instead of (\ref{nkj}), to the estimate
\be
\label{nkjexp}
n_k^j \sim \left[ \displaystyle\prod_{i = 1}^{j} \left(\frac{\tilde{k}}{a}\right)^2\,e^{X_k^{i-1}}\,e^{X_k^i} \right]^2 \, = \, \left(\frac{\tilde{k}}{a}\right)^{4 j}\,e^{2 X_k^0}\,e^{2 X_k^j}\,\displaystyle\prod_{i = 1}^{j-1} e^{4 X_k^i}
\ee
for the occupation number of the inflaton fluctuations after $j$ complete oscillations of the condensate. Indeed, during the $j^\mathrm{th}$ oscillation of the condensate, $n_k$ is first amplified by a factor of $e^{2 X_k^{j-1}}$ when the condensate rolls from $\phi = \phi_{j-1}$ to the minimum of the potential, and then by a factor of
$e^{2 X_k^{j}}\,\tilde{k}^4 / a^4$ when the condensate rolls back up to $\phi = \phi_j$. As before, the factor in
$\tilde{k}^4 / a^4 \ll 1$ results from the fact that the modes start to decrease exponentially when the condensate rolls back towards small values of $\phi$.

Note that, as in sub-section \ref{SubSecS}, Eq.~(\ref{nkjexp}) holds only for the modes which eventually increase exponentially when the condensate rolls back towards small values of $\phi$ (instead of decreasing exponentially during this whole stage), which dominate the second stage of preheating. These are the modes for which the factors in $e^{-X_k}$ in the transfer matrices (\ref{abaram}) and (\ref{baralphaa}) are negligible, i.e. the modes with
$\tilde{k} / a > e^{-X_k^j}$. Using (\ref{eXkj}) with $\tilde{k} / a \ll 1$, this condition reads
$\tilde{k} / a > \tilde{\phi}_j^{\sqrt{6}}$. Since $\tilde{\phi}_j$ in (\ref{phij}) is of the same order of magnitude for different values of $j \geq 1$, Eq.~(\ref{nkjexp}) holds for the modes with $\tilde{k} / a > \tilde{\phi}_1^{\sqrt{6}}$. On the other hand, Eq.~(\ref{nkjexp}) is not valid for the modes with $\tilde{k} / a < \tilde{\phi}_1^{\sqrt{6}}$, which decrease exponentially during all the time that the inflaton rolls back towards small values of $\phi$. These modes behave as the zero-mode $v_k \propto \dot{\phi}$ and oscillate without any net growth during the second stage of preheating. One can check that these include in particular the modes of the Hubble scale at the end of inflation, i.e. the modes with
$\tilde{k} / a \sim H_e / m_{\mathrm{min}} \sim \tilde{\phi}_0 \ll \tilde{\phi}_1^{\sqrt{6}}$ in rescaled
variables~\footnote{This is in contrast to the case without expansion of the universe, where we saw that only the modes with $\tilde{k} \ll \tilde{k}_0 \sim \tilde{\phi}_0$ decrease exponentially during all the time that the condensate rolls back towards small values of $\phi$. This is because in the absence of Hubble friction, the condensate rolls back to its initial value $\tilde{\phi}_0 \ll \tilde{\phi}_1$ after each oscillation. The second (exponentially growing) term in
(\ref{abar}) has then much more time to catch up with the first (exponentially decreasing) one, so that many more modes eventually increase exponentially when the condensate is rolling back.}. Indeed, we already saw in Section~\ref{Sec1} and Fig.~\ref{phink} that these modes follow the condensate velocity, $v_k \propto \dot{\phi}$.

In the following, instead of the occupation number, it will be more convenient to consider the variance per logarithmic momentum interval, $k^3\,|v_k|^2$. Note from (\ref{nk}) that $n_k \sim |v_k|^2\,k / a$. Then, using (\ref{nkjexp}) and
$\tilde{k} = 2 k / m_{\mathrm{min}}$, we have
\be
\label{k3v2j}
\frac{k^3}{a^3}\,|v_k^j|^2 \sim \frac{m_{\mathrm{min}}^2}{4}\,
\left(\frac{\tilde{k}}{a}\right)^{4 j + 2}\,e^{2 X_k^0}\,e^{2 X_k^j}\,\displaystyle\prod_{i = 1}^{j-1} e^{4 X_k^i}
\ee
after $j$ complete oscillations of the condensate. Because of the two different regimes (\ref{eXkj}) for each factor
in $e^{X_k^i}$ in (\ref{k3v2j}), the spectrum has a different shape in each interval
$\sqrt{3}\,\tilde{\phi}_{i-1} < \tilde{k} / a < \sqrt{3}\,\tilde{\phi}_{i}$. For instance, for
$\tilde{\phi}_1^{\sqrt{6}} < \tilde{k} / a < \sqrt{3}\,\tilde{\phi}_{1}$, all the factors in $e^{X_k^i}$ are constant, except $e^{X_k^0}$ (since we saw above that $\tilde{\phi}_0 \ll \tilde{\phi}_1^{\sqrt{6}}$). In that case, the spectrum increases as $k$ to the power $4 j + 2 - 2 \sqrt{6}$. On the other hand, for
$\tilde{k} / a > \sqrt{3}\,\tilde{\phi}_j$, all the factors in $e^{X_k^i}$ in (\ref{k3v2j}) decrease as
$\tilde{k}^{-\sqrt{6}}$. In that case, the spectrum decreases as $k$ to the power $-4 j (\sqrt{6} - 1) + 2$. In particular, the peak of the spectrum is located in between these two regimes, at $\tilde{k}_* / a \simeq \sqrt{3}\,\tilde{\phi}_{i}$ with $1 \leq i \leq j$. Using (\ref{phij}), this gives $\tilde{k}_* / a \approx \left(v / \mp\right)^{1/4}$. In terms of the un-rescaled momenta (\ref{tildes}), we have
\be
\label{kstar2}
\frac{k_*}{a} \approx \left(\frac{v}{\mp}\right)^{1/4}\,m_{\mathrm{min}}
\ee
for the characteristic physical momentum amplified during the second stage of preheating. Note that this is much larger than the characteristic momentum amplified during the first stage of preheating, $k_* / a \approx H_e$, so the spectrum is shifted towards the UV by a very large amount during the second stage.

From (\ref{k3v2j}), we can estimate the variance of the inflaton fluctuations $\psi_k = a^{-3/2}\,v_k$ after $j$ oscillations of the condensate as
\be
\langle \delta\phi^2 \rangle = \frac{1}{2\pi^2}\,\int \frac{dk}{k}\,k^3\,|\psi_k|^2 \sim
\frac{m_{\mathrm{min}}^2}{8\pi^2}\,\left(\frac{\tilde{k}_*}{a}\right)^{4 j + 2}\,e^{2 X_{k_*}^0}\,e^{2 X_{k_*}^j}\,\displaystyle\prod_{i = 1}^{j-1} e^{4 X_{k_*}^i}
\ee
where the RHS can be evaluated at the position of the peak, $\tilde{k}_* / a \approx \left(v / \mp\right)^{1/4}$. Using
(\ref{mmin}) and (\ref{eXkj}), this gives
\be
\label{dphi2j}
\langle \delta\phi^2 \rangle \sim \frac{M^4}{v^2}\,\left(\frac{v}{\mp}\right)^{-(\sqrt{6} - 1) j + 1/2} \ .
\ee
We can now estimate the number of oscillations of the condensate after which $\langle \delta\phi^2 \rangle \sim v^2$ and preheating ends. In (\ref{dphi2j}), $M$ depends on $v$ through the normalization of the CMB perturbations,
Eq.~(\ref{CMBMv}) with $p = 4$ for the model (\ref{potp4}). For the range $10^{-5} \lesssim v / \mp \lesssim 10^{-2}$ that we are interested in, the number of efolds before the end of inflation when cosmological scales leave the Hubble radius is of the order $N_* \approx 45$ - $50$. We then find that preheating in the model (\ref{potp4}) ends after only
\be
\label{numoscill}
j \approx \frac{10}{\mathrm{log}_{10}\left(\mp / v\right)}
\ee
oscillations of the condensate. This varies from $j = 2$ for $v / \mp = 10^{-5}$ to $j = 5$ for $v / \mp = 10^{-2}$. Thus,
for $v / \mp \lesssim 10^{-2}$, preheating is still very efficient despite the temporary exponential decrease of the modes.

\begin{figure}[htb]
\begin{center}
\includegraphics[width=7cm]{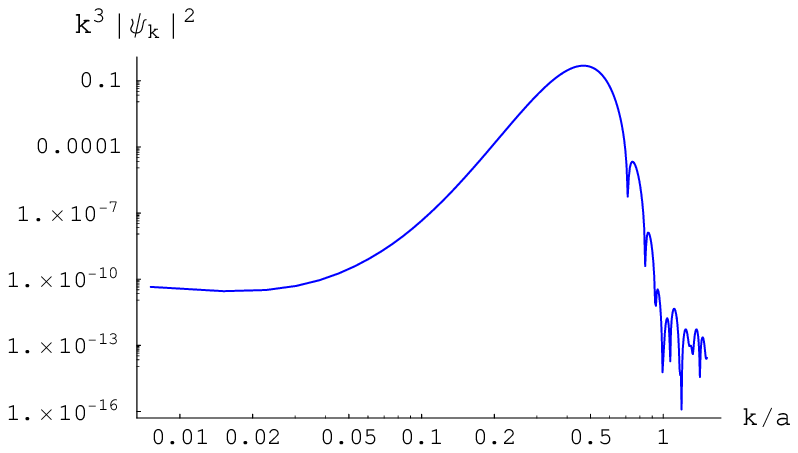} \hspace*{1cm}
\includegraphics[width=7cm]{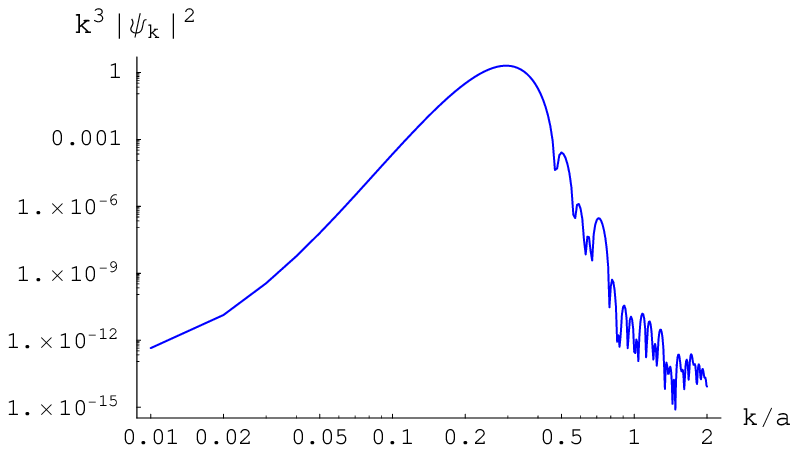} \vspace*{1cm}\\
\includegraphics[width=7cm]{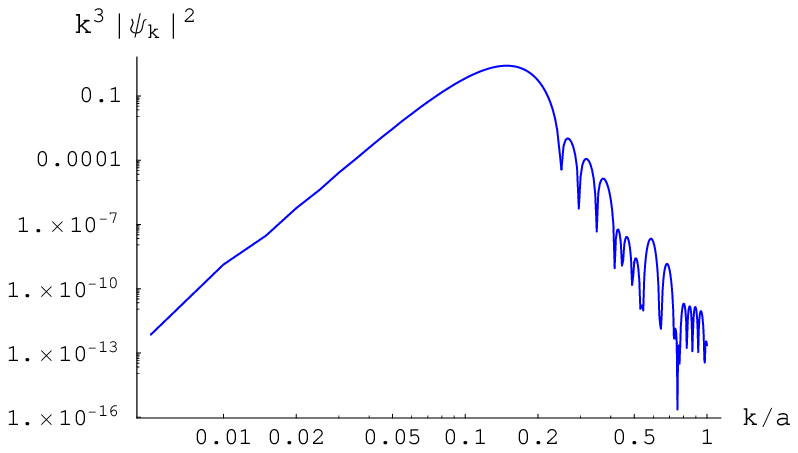} \hspace*{1cm}
\includegraphics[width=7cm]{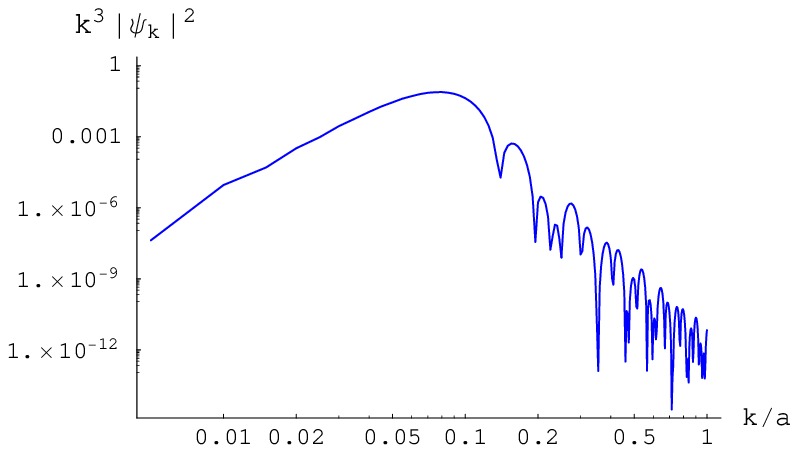}
\caption{Spectrum of inflaton fluctuations $k^3\,|\psi_k|^2$ at the end of the linear stage of preheating (when
$\langle \delta\phi^2 \rangle \approx v^2$) in the model (\ref{potp4}) for $v/\mp = 10^{-2}$ (top left), $10^{-3}$
(top right), $10^{-4}$ (bottom left) and $10^{-5}$ (bottom right). Preheating ends after $j \approx 6$, $4$, $3$ and $2$ oscillations of the inflaton condensate, respectively. The spectra are normalized in such a way that the peak amplitude is of order one at the end of preheating and they are shown as a function of the physical momentum $k/a$ in units of
$m_{\mathrm{min}}/2$.}
\label{SpecExp}
\end{center}
\end{figure}

The spectrum of the inflaton fluctuations calculated numerically at the end of the second stage of preheating is shown in Fig.~\ref{SpecExp} for different values of $v/\mp$. The results are in very good agreement with the analytical estimates (\ref{kstar2}) and (\ref{numoscill}). At momenta larger than the peak, some band structure remains due to destructive interferences, despite the expansion of the universe.

\subsection{Generalization to other models of small field inflation}
\label{SubSecGen}

The method that we developed above to study the second stage of preheating in the model (\ref{potp4}) is generic and can be applied to other models of small field inflation. To illustrate this, we now generalize our results to the models
(\ref{potp})
\be
\label{potpp}
V = M^4\,\left(1 - \frac{\phi^p}{v^p}\right)^2
\ee
with any $p > 2$. As discussed in Section \ref{SecBack}, $p > 2$ is the condition to have a small field inflation model. In this case, the inflaton mass at the minimum of the potential is given by
\be
\label{mminp}
m_{\mathrm{min}} = \sqrt{V''(v)} = \sqrt{2} p \, \frac{M^2}{v} \ .
\ee
The maximum of $-V''(\phi)$ is located at
\be
\label{phimp}
\phi_m = \left(\frac{p-2}{2 (2p-1)}\right)^{1/p}\,v
\ee
and the maximum momentum amplified by the tachyonic effect is
\be
\label{kmaxp}
\frac{k_{\mathrm{max}}}{a} = \sqrt{-V''(\phi_m)} = p\,\left(\frac{p-2}{2 (2p-1)}\right)^{\frac{p-2}{2p}}\,\frac{M^2}{v} \ .
\ee

As before, Hubble friction makes the second stage of preheating inefficient if $\phi_m \lesssim \phi_1$, where the amplitude of the condensate $\phi_1$ after one complete oscillation is now given by (\ref{phin}) with $j=1$ and
$\lambda = 2p$ for the model (\ref{potpp}). Comparing with (\ref{phimp}), we find that the condition (\ref{condnopre}) for the tachyonic effect to be inefficient remains approximately the same for any $p$. Similarly, the condition (\ref{cond1v}) for preheating to end during the first stage gives $v / \mp \lesssim 10^{-6} - 10^{-5}$ for any value of $p$. Thus, as before, the second stage of preheating in the model (\ref{potpp}) take place for
$10^{-5} \lesssim v / \mp \lesssim < 10^{-2}$.

\begin{figure}[htb]
\begin{center}
\includegraphics[width=7cm]{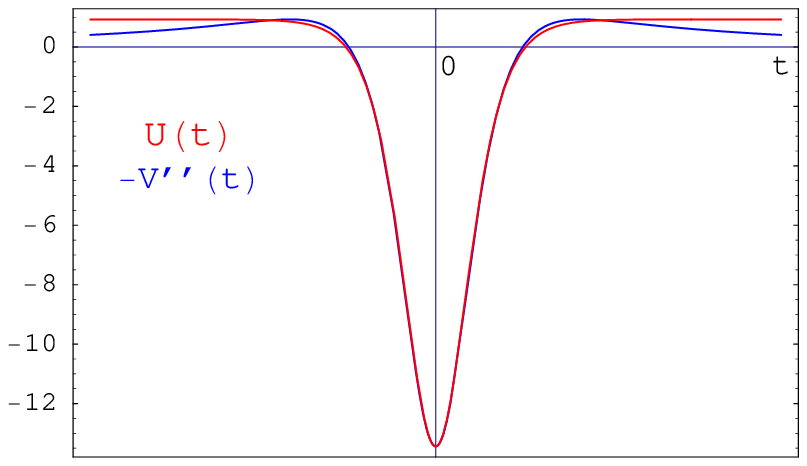}
\includegraphics[width=7cm]{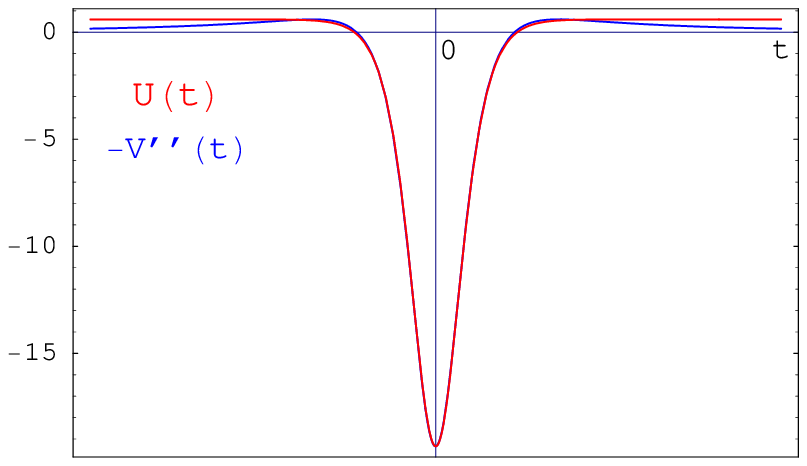}
\end{center}
\vspace*{-5mm}
\caption{Same as Fig.~\ref{Uapp}, but now for the model (\ref{potpp}) with $p=3$ (left) and $p=8$ (right).}
\label{Upotp}
\vspace*{-3mm}
\end{figure}

The second stage of preheating is again dominated by the regime of tachyonic oscillations for the modes with
$k \ll k_{\mathrm{max}}$. This can be studied analytically for any $p$ by a straightforward generalization of what we did in sub-sections~\ref{SubSecS} and \ref{SubSecExp}. The effective Schr\"{o}dinger potential $-\tilde{V}''(\phi)$ in the mode equation (\ref{modeq}) can again be approximated by the RHS of (\ref{Upot}) in the region $\phi > \phi_m$. This is illustrated in Fig.~\ref{Upotp} for $p=3$ and $p=8$. The parameters $k_{\mathrm{max}}$, $U_0$ and $\mu$ take different numerical values for different values of $p$, but they are still related by Eq.~(\ref{mumag}). This implies in particular that the WKB approximations (\ref{a}) and (\ref{abar}) are still related by the transfer matrix (\ref{abaram}), with the characteristic suppression in $k^2/k_{\mathrm{max}}^2 \ll 1$ due to the temporary exponential decrease of the modes when
$\dot{\phi} < 0$.

On the other hand, the rate of tachyonic growth $X_k$ depends on the value of $p$. Defining
\be
x_p = \frac{\sqrt{2 p (p-1)}}{p-2}
\ee
and proceding as in (\ref{eXkj}), we find the estimate
\be
\label{eXkjp}
e^{X_k^j} \sim \left\{\begin{array}{ccc}
\left(\frac{\tilde{k}}{a}\right)^{-x_p} & \mbox{ for } & \frac{\tilde{k}}{a} \gtrsim \tilde{\phi}_j^{(p-2)/2}
\vspace*{0.1cm} \\
\tilde{\phi}_j^{-x_p (p-2)/2} & \mbox{ for } & \frac{\tilde{k}}{a} \lesssim \tilde{\phi}_j^{(p-2)/2}
\end{array} \right.
\ee
for $\tilde{k}/a \ll 1$, where the "tildes" denote the rescaled variables (\ref{tildes}) and
\be
\label{phijp}
\tilde{\phi}_{j \geq 1} \approx \left(j\,\frac{v}{\mp}\right)^{1/p}
\ee
is the amplitude (\ref{phin}) of the condensate after $j \geq 1$ complete oscillations. As $p$ increases, for a fixed
ratio of $v/\mp$, the effect of Hubble friction on the amplitude of the condensate (\ref{phijp}) increases, so that the tachyonic growth of the low-momentum modes occurs on a shorter range of field values. However, the rate of growth of these modes increases also with $p$ and their net tachyonic growth $e^{X_k^j} \sim \left(\mp/v\right)^{\sqrt{(p-1)/(2p)}}$ increases slightly as $p$ increases.

We can then proceed as we did below Eq.~(\ref{k3v2j}). At the end of preheating, the peak of the spectrum of the inflaton fluctuations is now given by $\tilde{k}_* / a \approx \tilde{\phi}_1^{(p-2)/2}$. In terms of the physical variables, this gives
\be
\label{peakp2}
\frac{k_*}{a} \approx \left(\frac{v}{\mp}\right)^{\frac{p-2}{2p}}\,m_{\mathrm{min}} \ .
\ee
Therefore, in terms of $m_{\mathrm{min}}$, the peak of the spectrum moves towards the IR as $p$ increases. This is because, as we noted above, the rate of growth of the low-momentum modes increases when $p$ increases. The variance of the inflaton fluctuations grows as
\be
\langle \delta\phi^2 \rangle \sim \frac{M^4}{v^2}\,\left(\frac{v}{\mp}\right)^{\frac{p-2}{p}\left[-(x_p - 1) 2 j + 1\right]}
\ee
after $j$ complete oscillations of the condensate. As $p$ increases, the rate of growth of the fluctuations increases, but the ratio $M^4 / v^4$ fixed by the CMB normalization (\ref{CMBMv}) decreases. For any reasonable value of $p$ (say
$3 \leq p \leq 20$), we then find again that preheating ends in between $2$ and $6$ oscillations of the condensate for
$10^{-2} \lesssim v / \mp \lesssim 10^{-5}$.

\section{Other Non-Perturbative Decay Channels, Perturbative Decay and Reheat Temperature}
\label{SecPert}

We now study the decay of the inflaton into other fields, for a generic model (\ref{pot}, \ref{potor}) of small field inflation. We first consider other non-perturbative decay channels for the condensate and show that they are usually negligible with respect to the decay channel into inflaton fluctuations that we studied in the previous sections for
$v / \mp \lesssim 10^{-1}$. Preheating is then dominated by the mechanism that we studied above, leading to the very quick decay of the inflaton condensate into large and non-thermal fluctuations of itself. As usual, this is only the first stage of reheating. The inflaton fluctuations must then decay into other degrees of freedom and the universe must eventually thermalize. When the minimum of the inflaton potential $\phi = v$ corresponds to a fixed point of internal symmetries, the inflaton fluctuations may decay into massless fields and thermalization may be relatively fast. On the other hand, when
$\phi = v$ is not a fixed point of internal symmetries, the non-zero VEV of the inflaton at the minimum of the potential provides a large effective mass to the fields coupled to it. As a result, the inflaton can decay only into the fields to which it is weakly coupled, which therefore delays the decay. We will thus consider these two different cases separately in the following two sub-sections. In each case, we determine the reheat temperature and the resulting number of efolds $N_*$ before the end of inflation when cosmological perturbations leave the Hubble radius. We then determine, for different values of $p$, the explicit relation (\ref{CMBMv}) between $M$ and $v$ that follows from the normalization of the CMB anisotropies, and the region in the parameter space for which preheating ends in less than one oscillation of the condensate.

\subsection{Restored symmetries at the minimum}

Let us first consider the case where the minimum of the inflaton potential $\phi = v$ corresponds to a fixed point of internal symmetries. This is not a very frequent situation in small field models, but it arises for instance in the model of \cite{MSSM}. In that case, the inflaton can couple to massless fields at $\phi = v$. We can consider for example the interaction term
\be
\label{int1}
\frac{g^2}{2}\,\left(\phi - v\right)^2\,\chi^2
\ee
of the inflaton with another scalar field $\chi$, which is massless when $\phi = v$.

In that case, if the inflaton condensate did not decay during the first stage of preheating - i.e.
$v / \mp \gtrsim 10^{-6} - 10^{-5}$, see Section~\ref{Sec1} - it can lead to the non-perturbative production of
$\chi$-particles when it oscillates around the minimum $\phi = v$. If the $\chi$-particles are relatively long-lived, their number density can grow due to parametric resonance~\cite{KLS}. The condensate oscillates with an initial amplitude of the order of $\Phi \lesssim v$ and a frequency of the order of $1/m_{\mathrm{min}}$. The efficiency of the resonance is governed by the dimension-less parameter
\be
q \equiv \frac{g^2\,\Phi^2}{4\,m_{\mathrm{min}}^2} \propto g^2\,\frac{v^4}{M^4}
\ee
where we used $m_{\mathrm{min}} \sim M^2 / v$. The relation (\ref{CMBMv}) shows that the resonance can be broad ($q > 1$) or narrow ($q < 1$), depending on the value of $g^2$ and $p$. For $p = 3$ or $4$, we have usually $q \gg 1$, which is the most efficient case. In that case, the occupation number of the $\chi$-particles grows as $e^{2 \mu_k j}$ after $j$ oscillations of the inflaton condensate, where the maximal value of $\mu_k$ is $\mu_k \simeq 0.28$~\cite{KLS}. This growth is clearly negligible with respect to the mechanism of tachyonic oscillations that we studied in
sub-section~\ref{SubSecGen}.

Another possible mechanism for preheating is called instant preheating~\cite{instant}. This may occur if $\chi$-particles are produced non-adiabatically when $\phi = v$ and then decay into other particles in less than one oscillation of the inflaton condensate. Again, this requires that the condensate did not decay before reaching the minimum of the potential, $v / \mp \gtrsim 10^{-6} - 10^{-5}$. When it crosses the minimum, it produces massless $\chi$-particles with a number
density~\cite{instant}
\be
\label{nchi}
n_\chi \approx \frac{\left(g\,\dot{\phi}_v\right)^{3/2}}{8\,\pi^3}
\ee
via the interaction (\ref{int1}). Here $\dot{\phi}_v \sim M^2$ is the velocity of the condensate at the minimum
$\phi = v$. After they have been produced, the effective mass of the $\chi$-particles increases as
$m_\chi = g\,|\phi - v|$. If these $\chi$-particles decay, for instance into fermions with an interaction term
\be
h\,\chi\,\psi\,\bar{\psi} \, ,
\ee
then their decay rate is given by
\be
\Gamma_{\chi \rightarrow \psi\,\bar{\psi}} = \frac{h^2\,m_\chi}{8\pi} = \frac{h^2\,g\,|\phi - v|}{8 \pi}
\ee
which is maximum when $|\phi - v| \sim v$ is maximum. For the mechanism of instant preheating to be efficient, this decay should occur in less than one oscillation of the inflaton condensate, i.e. $\Gamma \gtrsim m_{\mathrm{min}}$. This condition gives
\be
\label{condinst}
h^2 \gtrsim \frac{8 \pi\,m_{\mathrm{min}}}{g\,|\phi - v|} \gtrsim \frac{10^2}{g}\,\frac{M^2}{v^2}
\ee
where we have used $m_{\mathrm{min}} \sim M^2 / v$ and $|\phi - v| \lesssim v$. The condition (\ref{condinst}) may be easily satisfied for $p = 3$ or $4$, see Eq.~(\ref{CMBMv}).

In that case, once the $\chi$-particles have been produced when $\phi = v$, they decay after a time $\Delta t \sim 1 / \Gamma$, when the value of the inflaton condensate is
$\phi_d - v \approx \Delta t\,\dot{\phi}_v$. This gives
\be
|\phi_d - v| \approx \left(\frac{8\pi\,\dot{\phi}_v}{h^2\,g}\right)^{1/2}
\ee
when the $\chi$-particles decay. The decay produces $\psi$-particles with a typical energy
$E_\psi = m_\chi / 2$ where $m_\chi = g\,|\phi_d - v|$. The energy density of the $\psi$-particles
$\rho_\psi = n_\psi\,E_\psi$ compared to the energy density of the inflaton condensate $\rho_\phi = \dot{\phi}_v^2 / 2$
is then given by
\be
\frac{\rho_\psi}{\rho_\phi} = \frac{n_\chi\,m_\chi}{\dot{\phi}_v^2 / 2} \approx 0.05\,\frac{g^2}{h}
\ee
where we have used Eq.~(\ref{nchi}). Thus this mechanism can be very efficient for suitable interactions between the different fields. For $v / \mp \gtrsim 10^{-6} - 10^{-5}$ and $h \ll g^2$, it can compete with the tachyonic decay into inflaton fluctuations if the minimum $\phi = v$ is indeed a fixed point of internal symmetries. The last condition is necessary because otherwise the non-adiabatic production of $\chi$-particles is negligible from the beginning, see the next sub-section.
A particularly interesting realisation of the instant preheating framework has been recently provided in \cite{anum} where inflation occurs along  the $LLe$  D-flat direction in the MSSM. After the end of inflation, the inflaton oscillates around a point of enhanced gauge symmetry. Tachyonic preheating would end in a large ($\approx 10$) oscillations while in fact instant preheating implies that the inflaton loses enough energy in the first oscillation never to go back to the tachyonic region. In the subsequent oscillations, instant preheating transfers the energy from the inflaton to the SM quarks, leptons and their supersymmetric partners.

Note also that, even in the more generic case where the dominant mechanism for preheating is the tachyonic decay into inflaton fluctuations, fluctuations of other fields may then be quickly amplified too via their interactions with the inflaton fluctuations~\cite{equilibrium}. Thus in general, at the end of preheating, the universe is dominated by large fluctuations of the inflaton and other fields coupled to it. Since the inflaton fluctuations are massive, they become quickly non-relativistic and dominate the total energy density before they decay. This subsequent decay of the inflaton can be described by the perturbative theory of reheating~\cite{KLS}. We now consider this stage of perturbative decay.

When $\phi = v$ is a fixed point of internal symmetries, the inflaton can couple to massless fields and quickly decay into these. Trilinear interactions allow for the inflaton to decay completely. This is provided for instance by an interaction term
\be
h\,|\phi - v|\,\psi\,\bar{\psi}
\ee
with fermions $\psi$, which are again massless for $\phi = v$. The decay rate of the inflaton fluctuations
$\delta\phi = \phi - v$, whose effective mass is~\footnote{When the inflaton fluctuations are large,
$\langle \delta\phi^2 \rangle \sim v^2$, they contribute to their own effective mass which may then differ from
$\sqrt{V''(v)}$. This however does not change the order of magnitude.} $m_\phi \approx m_{\mathrm{min}}$, then reads
\be
\Gamma_{\phi \rightarrow \psi\,\bar{\psi}} \approx \frac{h^2\,m_{\mathrm{min}}}{8\pi} \ .
\ee
In small field models, $m_{\mathrm{min}} \sim M^2 / v$ can be much larger than the Hubble rate during inflation,
$H_e \sim M^2 / \mp$. If the coupling constant $h$ is not too small, $\Gamma > H_e$ and the inflaton then decays in less than one Hubble time after the end of inflation. Assuming that the decay products quickly thermalize, the reheat temperature is then almost of the order of the energy scale during inflation
\be
\label{InstReh}
T_R \sim M \ .
\ee
We will call this case "instantaneous reheating". We will see that the situation can be very different when $\phi = v$ is not a fixed point of internal symmetries. Note also that Eq.~(\ref{InstReh}) implies that inflation should occur at a rather low energy scale in order to satisfy the usual upper bound on $T_R$ to avoid the thermal over-production of gravitinos, $M \lesssim 10^9$ GeV or so.

\begin{figure}[htb]
\begin{center}
\includegraphics[width=11cm]{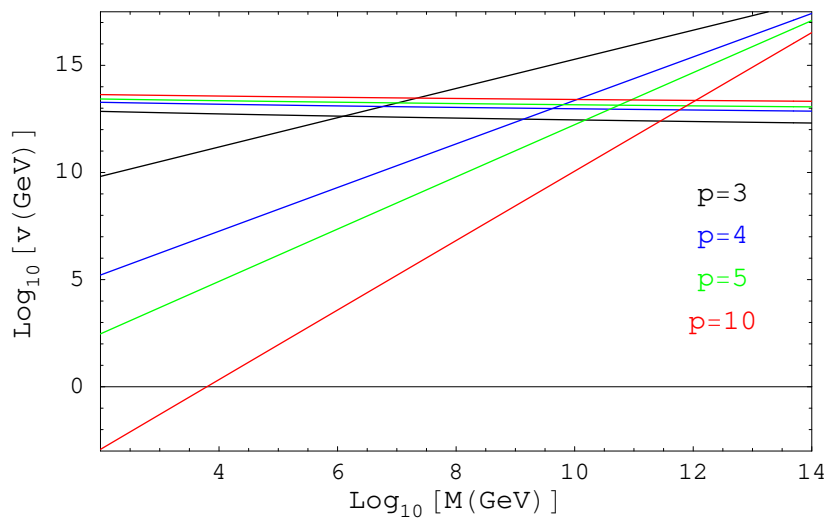}
\caption{Relation between $v$ and $M$ from CMB normalization, for $p = 3$ (black), $4$ (blue), $5$ (green) and $10$ (red), in the case of "instantaneous reheating". The almost horizontal lines that are roughly superimposed at the top of the plot show the upper bound (\ref{cond1v}) on $v$ below which preheating occurs in less than one oscillation of the condensate, for the same values of $p$.}
\label{vMInst}
\end{center}
\end{figure}

Let us now determine the explicit relation between $M$ and $v$ from the normalization of the CMB anisotropies (\ref{CMBMv}) in the case of instantaneous reheating. In general, the number of efolds $N_*$ before the end of inflation when observable scales leave the Hubble radius is given by
\be
\label{Nstar}
N_* \simeq 63.9 + \ln\left(\frac{V_{\mathrm{infl}}^{1/4}}{M_{\mathrm{infl}}}\right) +
\frac{1}{12}\,\ln\left(\frac{\rho_{\mathrm{reh}}}{V_{\mathrm{infl}}}\right)
\ee
where $M_{\mathrm{infl}} \simeq 6.6\,10^{16}$ GeV, see (\ref{CMB}). Here $V_{\mathrm{infl}} = M^4$ is the energy density at the end of inflation and $\rho_{\mathrm{reh}}$ the energy density when the universe becomes radiation-dominated. When reheating is instantaneous, $\rho_{\mathrm{reh}} = V_{\mathrm{infl}}$ so that
\be
\label{Ninst}
N_* \simeq 63.9 + \ln\left(\frac{M}{M_{\mathrm{infl}}}\right) \ .
\ee
Inserting this into (\ref{CMBMv}) gives the explicit relation between $v$ and $M$. This is shown in Fig.~\ref{vMInst} for
$\lambda = 2p$ and different values of $p$. The almost horizontal lines that are roughly superimposed at the top of the plot show the upper bound (\ref{cond1v}) on $v$ below which preheating occurs in less than one oscillation of the condensate, for the same values of $p$. For $p = 3$, preheating occurs in less than one oscillation of the condensate only for small energy scales during inflation, $M \lesssim 10^6$ GeV. This upper bound increases with $p$, e.g. $M \lesssim 10^9$ GeV for $p = 4$, which is similar to the gravitino bound, and $M \lesssim 10^{11}$ GeV for $p = 5$, which corresponds to the intermediate scale of SUSY breaking in gravity-mediated scenarios.

\subsection{Broken symmetries at the minimum}

We now consider the case where $\phi = v$ is not a fixed point of internal symmetries. We will first consider the decay of the inflaton into fermions $\psi$ and another scalar field $\chi$ via the renormalizable interaction terms
\be
\label{couplings}
h\,\phi\,\psi\,\bar{\psi} \hspace*{0.5cm} , \hspace*{0.5cm} \frac{g^2}{2}\,\phi^2\,\chi^2 \,.
\ee
These interactions arise naturally if $\phi = 0$ is a fixed point of internal symmetries, which may help to ensure the flatness of the inflaton potential during inflation. At the minimum of the inflaton potential, the non-zero VEV $v$ of the inflaton provides an effective mass to $\psi$ and $\chi$ via the interactions (\ref{couplings}), $m_\psi = h\,v$ and
$m_\chi = g\,v$.

As in the previous sub-section, let us first check if the interactions (\ref{couplings}) can lead to non-perturbative decay channels for the inflaton condensate that could compete with the decay into inflaton fluctuations that we studied in Sections~\ref{Sec1} and \ref{Sec2}. We will consider the non-perturbative production of the scalar field $\chi$ since the non-perturbative production of fermions would be similar but limited by Pauli blocking. Let us first suppose that any bare mass of the $\chi$-field is negligible compared to the contribution coming from the interaction (\ref{couplings}), so that the effective mass of $\chi$ is simply $m_\chi = g\,\phi$. Efficient production of $\chi$-particles may occur when the effective mass varies non-adiabatically with time, i.e.
\be
\label{dotmm2}
\frac{|\dot{m}_\chi|}{m_\chi^2} = \frac{|\dot{\phi}|}{g\,\phi^2} \gtrsim 1 \ .
\ee
From Eq.~(\ref{Econs}), we have $\dot{\phi} \propto \phi^{p/2}$ for $\phi \ll v$. This implies that
$\dot{m}_\chi / m_\chi^2$ is maximum at small field values for $p \leq 4$ and at $\phi = v$ for $p > 4$.
At the beginning of preheating, $\phi = \phi_e$, Eq.~(\ref{dotmm2}) gives
\be
\left.\frac{\dot{m}_\chi}{m_\chi^2}\right|_{\phi = \phi_e} = \frac{1}{g}\,
\frac{\left[p-1 + (p-2) N_*\right]^{-(p-1)/(p-2)}}{\sqrt{6}\,\left(p-1\right)^{(p-3)/(p-1)}}\,
\frac{M_{\mathrm{infl}}^2}{\mp^2}
\ee
where we have used Eqs.~(\ref{phiend}), (\ref{dotphiend}) and (\ref{CMBMv}). This shows that $\dot{m}_\chi \ll m_\chi^2$ at small field values, except for very small values of the coupling constant $g$. However, for such small values of $g$, the contribution $g\,\phi$ to the effective mass of $\chi$ is smaller than the Hubble rate during inflation. In order for
$\chi$ to remain massive during inflation, we must include a bare mass $\bar{m}_\chi > H$, which in that case is not anymore negligible with respect to $g\,\phi$. When this is done, we find that $\dot{m}_\chi \ll m_\chi^2$ at
$\phi = \phi_e$ for any value of $g$. Therefore, there is no non-adiabatic production of $\chi$-particles for $p \leq 4$.

For $p > 4$, $\dot{m}_\chi / m_\chi^2$ is instead maximum at the minimum of the potential, $\phi = v$. We have
$\dot{\phi} \simeq \sqrt{2}\,M^2$ there, so Eq.~(\ref{dotmm2}) gives
\be
\left.\frac{\dot{m}_\chi}{m_\chi^2}\right|_{\phi = v} \approx \frac{M^2}{g\,v^2} \ .
\ee
For $g \sim 1$, this is of the order of $M^2 / v^2$ which, from Eq.~(\ref{CMBMv}), can be larger than one only for
$p > 4$ and $v$ much smaller than $\mp$. However, as discussed in Section~\ref{Sec1}, for such small values of $v$ the inflaton condensate decays before it reaches the minimum of the potential, so that there is no non-adiabatic production of
$\chi$-particles at $\phi = v$ in that case either. Therefore, the only case where such a non-adiabatic production is possible if for $p > 4$ and $g \ll 1$, where $m_\chi$ may vary non-adiabatically around $\phi = v$ if the inflaton condensate did not decay during the first stage of preheating. As in the previous sub-section, either parametric resonance or instant preheating are a priori possible in this special case. However, the condition (\ref{condinst}) is not satisfied in that case, so the $\chi$-particles can only decay after several oscillations of the condensate and instant preheating does not take place. As for the production of $\chi$-particles by parametric resonance, it is again negligible with respect to the tachyonic amplification of the inflaton fluctuations. We therefore conclude that, when $\phi = v$ is not a fixed point of internal symmetries, preheating is in general completely dominated by the tachyonic effect that we studied in Sections~\ref{Sec1} and \ref{Sec2}.

As in the previous sub-section, we now turn to the perturbative decay of the inflaton after preheating. Since we consider inflaton fluctuations around the minimum, $\phi = v + \delta\phi$, the interaction term $g^2\,\phi^2\,\chi^2$ with the scalar field $\chi$ now includes a trilinear coupling $g^2\,v\,\delta\phi\,\chi$. Thus both interactions in
(\ref{couplings}) allow for the complete decay of the inflaton. The corresponding decay rates are given by
\be
\label{decrates}
\Gamma_{\phi \rightarrow \psi\bar{\psi}} = \frac{h^2\,m_\phi}{8\pi} \hspace*{0.5cm} , \hspace*{0.5cm}
\Gamma_{\phi \rightarrow \chi\chi} = \frac{g^4\,v^2}{8\pi\,m_\phi}
\ee
when the fermion $\psi$ and the scalar $\chi$ are much lighter than the inflaton $\phi$, and when $H \ll m_\phi$ (which is always satisfied for $H \sim \Gamma$, when most of the decay occurs). The mass of the inflaton is again of the order of the curvature of the bare potential at the minimum, $m_\phi \approx m_{\mathrm{min}} \sim M^2 / v$. On the other hand, as we already mentioned, when $\phi = v$ is not a fixed point of internal symmetries, the interactions (\ref{couplings}) provide an effective mass to $\psi$ and $\chi$ at $\phi = v$, $m_\psi = h\,v$ and $m_\chi = g\,v$. However, the decay of the inflaton is kinematically possible only if its mass is larger than at least twice the mass of the decay products,
$m_\phi/2 > m_\psi, m_\chi$. This requires
\be
h, g < \frac{m_\phi}{2\,v}
\ee
so that Eqs.~(\ref{decrates}) give
\be
\label{decbound}
\Gamma_{\phi \rightarrow \psi\bar{\psi}} < \frac{m_{\phi}^3}{32 \pi\,v^2}  \hspace*{0.5cm} , \hspace*{0.5cm}
\Gamma_{\phi \rightarrow \chi\chi} < \frac{m_{\phi}^3}{128 \pi\,v^2} \ .
\ee
Thus the decay is possible only for sufficiently small coupling constants, which in turn delays the decay.

This kinematical blocking of the inflaton decay when $\phi = v$ is not a fixed point of internal symmetries occurs because the non-zero VEV $v$ of the inflaton at the minimum provides a large effective mass to the fields coupled to it. On the other hand, if some fields couple to the inflaton through derivative interactions only, these fields may remain massless despite the non-zero VEV of the inflaton. One may therefore wonder if derivative interactions could provide the fastest decay channel for the inflaton. The inflaton may have such interactions for instance with a gauge field $A_\mu$, with the trilinear couplings
\be
\label{phigauge}
\frac{\phi}{4\,\Lambda}\,F_{\mu \nu}\,F^{\mu \nu} \hspace*{0.5cm} , \hspace*{0.5cm}
\frac{\phi}{4\,\Lambda}\,F_{\mu \nu}\,\tilde{F}^{\mu \nu}
\ee
where $F_{\mu \nu}$ is the gauge field strength,
$\tilde{F}^{\mu \nu} = \epsilon^{\mu \nu \rho \sigma}\,F_{\rho \sigma} / 2$ is the dual field strength and $\Lambda$ is a mass scale such as the cutoff scale or the symmetry breaking scale $v$. Notice that the coupling involving the dual field strength implies that $\phi$ should be a pseudo-scalar and therefore the inflaton potential should be an even function. The first and second couplings in (\ref{phigauge}) are natural if the inflaton $\phi$ is a modulus or an axion, respectively. The corresponding decay rates are given by
\be
\label{decAbound}
\Gamma_{\phi \rightarrow A A} = \frac{m_\phi^3}{64 \pi\,\Lambda^2} \, \leq \, \frac{m_\phi^3}{64 \pi\,v^2}
\ee
where, in the last inequality, we have taken the condition $\Lambda \geq v$. When the interactions (\ref{phigauge}) arise as the leading term in a series expansion in powers of $\phi$, this condition should be satisfied for the series to be in a controllable regime. For instance, if $\Lambda$ is the cutoff scale of the effective theory, $v < \Lambda$ is required for the model of small field inflation to lie within the regime of validity of the effective theory. Even if we had
$v \gg \Lambda$, the first (modular) coupling in (\ref{phigauge}) with $\phi = v + \delta \phi$ includes a contribution
$v/(4\Lambda) F_{\mu \nu}\,F^{\mu \nu}$ to the kinetic term of the gauge field. Rescaling to the canonical gauge field then leads to the same bound (\ref{decAbound}) for the decay rate of $\delta \phi$. This does not arise for the second (axionic) coupling in (\ref{phigauge}) because $F_{\mu \nu}\,\tilde{F}^{\mu \nu}$ is a total derivative. One could then imagine a situation where the inflaton has only such an axionic coupling and the decay rate would be greater than the bound in
(\ref{decAbound}). However, as we just argued, this is very unlikely.

Interestingly, the upper bounds (\ref{decbound}, \ref{decAbound}) on the decay rates for each interaction are all of the same order of magnitude. In the following, we will assume the presence of at least one such interaction with a coupling constant that maximizes the decay rate. Using $m_\phi \approx m_{\mathrm{min}} \approx\,M^2 / v$
(see e.g. (\ref{mminp}) for the model (\ref{potpp})), we then have
\be
\label{GaDel}
\Gamma \approx \frac{M^6}{v^5}
\ee
for the decay rate of the inflaton. Compared to the Hubble rate at the end of inflation $H_e = M^2 / (\sqrt{3} \mp)$, this gives
\be
\label{GaHe}
\frac{\Gamma}{H_e} \approx \frac{M^4}{v^4}\,\frac{\mp}{v} \ .
\ee
Eq.~(\ref{CMBMv}) shows that $\Gamma \ll H_e$ in large regions of the parameter space. In that case, the universe enters a long matter-dominated stage after preheating, until the non-relativistic inflaton fluctuations decay completely.

\begin{figure}[htb]
\begin{tabular}{cc}
\begin{minipage}[t]{8.5cm}
\begin{center}
\includegraphics[width=8.5cm]{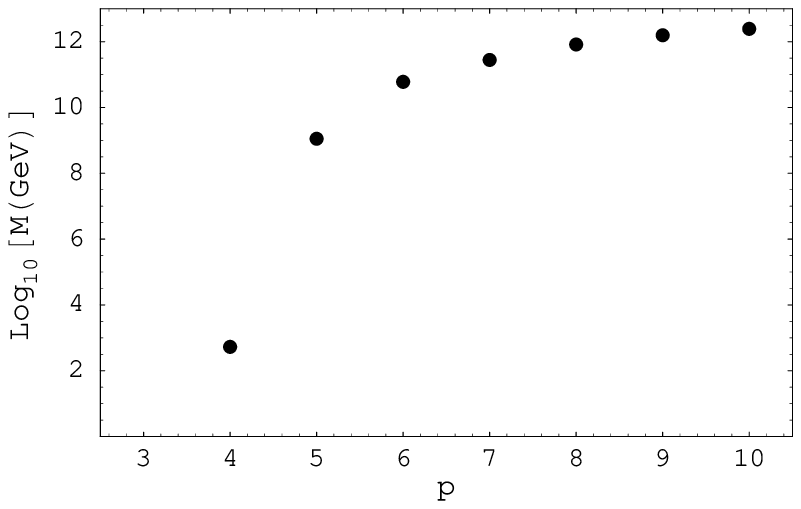}
\caption{Maximum value of $M$ (in GeV) for which the condition (\ref{CondReInst}) is satisfied, as a function of
$p$, when $\phi = v$ is not a fixed point of internal symmetries. Reheating is "instantaneous" for
$p > 10/3$ if $M$ is smaller than the value indicated on the plot.}
\label{MmaxReInst}
\end{center}
\end{minipage}&
\hspace*{0.5cm}
\begin{minipage}[t]{8.5cm}
\begin{center}
\includegraphics[width=8.5cm]{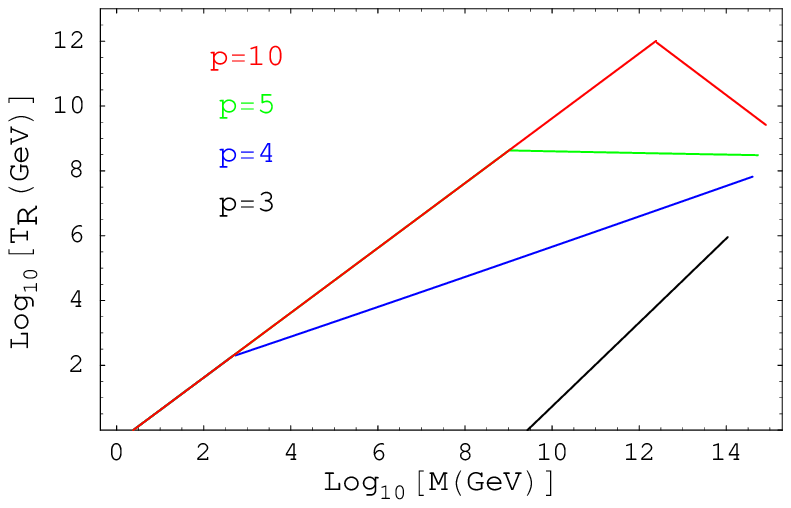}
\caption{Reheat temperature as a function of the energy scale during inflation, for $p = 3$ (black), $4$ (blue), $5$ (green) and $10$ (red), when $\phi = v$ is not a fixed point of internal symmetries. See main text for details.}
\label{TRM}
\end{center}
\end{minipage}
\end{tabular}
\end{figure}

Let us first determine the conditions on the parameters for which, despite the suppression of the decay rate, the inflaton can nevertheless decay in less than one Hubble time after the end of inflation. This occurs if $\Gamma > H_e$. Using
(\ref{CMBv}) in (\ref{GaHe}), this gives
\be
\label{CondReInst}
C^{-5(p-2)/p}\,\frac{M_{\mathrm{infl}}^4}{\mp^4}\,
\left(\frac{M}{M_{\mathrm{infl}}}\right)^{-2(3p-10)/p} \, > \, 1 \ .
\ee
where $C$ is defined in (\ref{defC}). When this condition is satisfied (and assuming again that the inflaton decay products quickly thermalize), reheating is "instantaneous": $T_R \sim M$ and the number of efolds is given by
(\ref{Ninst}). As discussed below Eq.~(\ref{CMBv}), we have usually $C > 1$ and
$M < M_{\mathrm{infl}}$. This implies that the condition (\ref{CondReInst}) cannot be satisfied for
$p \leq 10/3 \simeq 3.33$. Thus reheating is never instantaneous for $p = 3$. For larger values of $p$, the condition
(\ref{MmaxReInst}) with (\ref{defC}, \ref{Ninst}) gives an upper bound on $M$ below which reheating is instantaneous. This upper bound on $M$ is shown in Fig.~\ref{MmaxReInst} for different values of $p$. For $p = 4$, reheating is instantaneous only if $T_R \sim M \lesssim 500$ GeV, which is very small but can still be compatible with electroweak baryogenesis. For larger values of $p$, reheating is instantaneous for larger values of $M$, e.g. $M \lesssim 10^{11}$ GeV for $p=6$.

Let us now consider the other case, when $\Gamma < H_e$. In that case, reheating is delayed until the Hubble rate $H$ drops below $\Gamma$ and the reheat temperature is then given by $T_R \approx 0.5\,\sqrt{\Gamma\,\mp}$. Using (\ref{GaDel}),
this gives
\be
\label{TRDel}
T_R \, \approx \, \frac{M^2}{v^2}\,\left(\frac{\mp}{v}\right)^{1/2} \, = \, C^{-\frac{5(p-2)}{2p}}\,
\frac{M_{\mathrm{infl}}^2}{\mp^2}\,\left(\frac{M}{M_{\mathrm{infl}}}\right)^{-(3p-10)/p}\,M
\ee
where we have used (\ref{CMBv}) in the second equality. The number of efolds is now given by (\ref{Nstar}) with
$\rho_{\mathrm{reh}} = 3 \mp^2\,\Gamma^2$. Using (\ref{GaDel}, \ref{CMBv}), this gives
\be
N_* + \frac{5(p-2)}{6p}\,\ln C \approx 63.9 + \frac{10}{3p}\,\ln\left(\frac{M}{M_{\mathrm{infl}}}\right)
\ee
From this, we can calculate $N_*$ and $C$ as a function of $M$ and then use (\ref{TRDel}) to calculate $T_R$ as a function of $M$. The result is shown in Fig.~\ref{TRM} for different values of $p$. The curves for different values of $p > 10/3$ are superposed for sufficiently low $M$, because reheating is then instantaneous and $T_R \simeq 0.5\,M$ independently of
$p$. Then, when $M$ becomes greater than a certain value that depends on $p$ (the value shown in Fig.~\ref{MmaxReInst}), reheating starts to be delayed and $T_R$ increases with $M$ more slowly than $T_R \propto M$. In fact, Eq.~(\ref{TRDel}) shows that, for $p > 5$, the reheat temperature even decreases when the energy scale of inflation increases! This is because, in that case, the requirement that the perturbative decay of the inflaton is kinematically possible leads to a more stringent upper bound on $\Gamma$ when $M$ increases, so that reheating takes a longer time to complete. Meanwhile, for $p = 3$, we saw above that we have $\Gamma \ll H_e$ so that the reheat temperature is typically much smaller than $M$.

\begin{figure}[htb]
\begin{center}
\includegraphics[width=11cm]{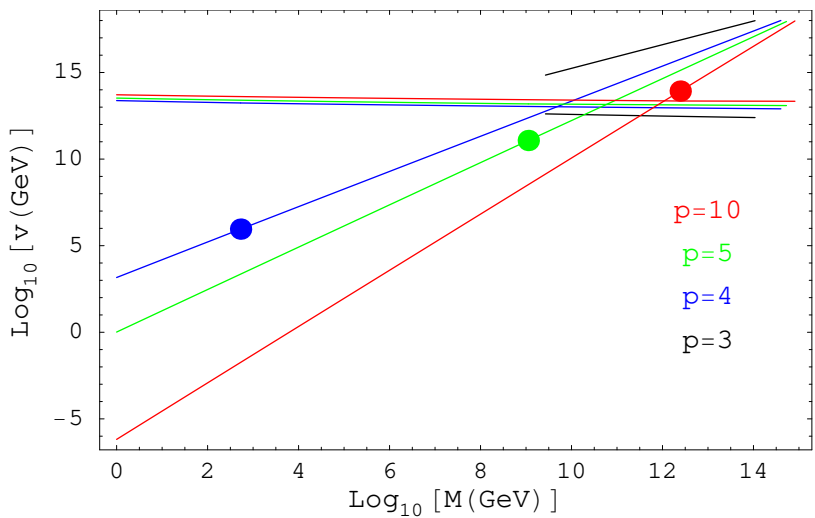}
\caption{Relation between $v$ and $M$ from CMB normalization, for $p = 3$ (black), $4$ (blue), $5$ (green) and $10$ (red), when $\phi = v$ is not a fixed point of internal symmetries. Only the values for which $T_R >$ GeV and $v < \mp$ are shown. For $p = 4$, $5$ and $10$, the dots indicate the values of $v$ and $M$ beyond which reheating is instantaneous. The almost horizontal lines that are roughly superimposed at the top of the plot show the upper bound (\ref{cond1v}) on $v$ below which preheating occurs in less than one oscillation of the condensate, for each value of $p$.}
\label{vGMG}
\end{center}
\end{figure}

Finally, we show in Fig.~\ref{vGMG} the relation (\ref{CMBv}) between $v$ and $M$ for different values of $p$. In each case, we only show the values for which $v < \mp$ and $T_R >$ GeV, and the dot indicates the values of $M$ and $v$ below which reheating is instantaneous. The almost horizontal lines that are roughly superimposed at the top of the plot show the upper bound (\ref{cond1v}) on $v$ below which preheating occurs in less than one oscillation of the condensate, for different values of $p$. For $p = 3$, since the decay rate is very small when $\phi = v$ is not a fixed point of internal symmetries, preheating must occur in more than one oscillation for the universe to reheat before BBN. For the other values of $p$, all the cases are possible depending on the energy scale of inflation: preheating in less or more than one oscillation and instantaneous or delayed reheating.
Each case will lead to different predictions for the cosmological consequences of preheating after small field inflation, such as the production of primordial
black holes and gravitational waves.

%%%%%%%%%%%%%%%%%%%%%%%%%%%%%%%%%%%%%%%%%%%%%%%%%%%%%%%%%%%%%%%%%%%%%%%%%%%%%%%%%

\section{Conclusion}
\label{SecConclu}

We have studied the linear stage of preheating in the class of small field inflation models, where the curvature of the
inflaton potential is negative during inflation. Although this is one of the most common classes of inflationary models,
preheating after small field inflation remained much less studied than preheating after chaotic and hybrid inflation.
On the analytical side, the problem is complicated by the fact that closed form solutions for the background evolution
of the inflaton condensate are in general not available, while standard approximation like WKB for the evolution of
perturbations are not accurate. Nevertheless, we saw that a detailed analytical description of preheating in this
class of models is possible. The analytical methods that we developed in this paper may also be applied to the study of
preheating in more complicated models.

We showed that preheating after small field inflation is usually dominated by the tachyonic amplification of inflaton
fluctuations in the intervals of time when the inflaton condensate rolls in the region where the curvature of its
potential is negative. A peculiar feature of this process is that the inflaton fluctuations experience a succession of
exponential growths and decreases, so we called it "tachyonic oscillation". The exponential decreases of the fluctuations
in these intervals of time arise because the coefficient of the decreasing mode is much larger than the coefficient of the
growing mode, by an amount that depends on the scale of the fluctuation. Despite this temporary exponential decrease of
the fluctuations, the full process is very efficient and we showed that preheating completes typically after less than
five oscillations of the inflaton condensate. The range of scales of the fluctuations amplified in the course of
preheating is very wide, extending from the Hubble scale to the curvature of the potential at the minimum, with a peak
given by Eq.~(\ref{peakp2}) when preheating completes in more than one oscillation of the condensate.

When the condition (\ref{cond1v}) is satisfied, the first tachyonic instability is so efficient that preheating completes
in less than one oscillation of the condensate. In that case, the spectrum of the inflaton fluctuations at the end of
preheating is peaked around the Hubble scale, or even slightly outside the Hubble radius because of the residual amount
of inflation during preheating. Density fluctuations at the Hubble scale may then lead to an abundant production of primordial
black holes, see also~\cite{kudoh2}, which can put constraints on small field inflation models, see e.g.~\cite{PBH} for a recent
update on observational constraints on primodrial black holes. The large field fluctuations
amplified by preheating lead also to the production of gravitational waves (GW), with a peak frequency today~\cite{GWpre}
\be
\label{fstarGW}
f_* \approx \left(\frac{k_* / a}{H_p}\right)\,\left(\frac{\rho_p^{1/4}}{10^{11}\,\mathrm{GeV}}\right)\,10^3\,\mathrm{Hz}
\ee
where $H_p$ and $\rho_p$ are the Hubble rate and the energy density during preheating, and $k_*/a$ is the charactersitic
momentum amplified by preheating~\footnote{Eq.~(\ref{fstarGW}) is valid when the universe becomes quickly
radiation-dominated after preheating. Note also that, in models where gauge fields play an important role in the dynamics
of preheating, extra peaks may appear at well-distinct frequencies in the final GW spectrum~\cite{GWvec}.}. GW from
preheating after chaotic and hybrid inflation tend to have a frequency today above $10^3$ Hz, which is too high to be
observable by high-sensitivity interferometric experiments, either because $\rho_p^{1/4} \gg 10^{11}$ GeV or because
$k_* / a \gg H_p$~\cite{GWpre}. GW from the non-perturbative decay of condensates different from the inflaton, in
particular super-symmetric flat directions, are more promising in this respect although the GW amplitude may be suppressed
in that cases~\cite{GWflat}. On the other hand, GW from preheating after small field inflation may fall naturally in
the frequency range accessible by ground-based and even space-based interferometers when $k_* / a \sim H_p$ and the
energy scale of inflation is small enough, which is common in this class of models. We also saw that preheating after
small field inflation may be followed by a long matter-dominated stage before the universe thermalizes, which would
further redshift and dilute these GW. A detailed study of the production of primordial black holes and GW from preheating
after small field inflation is currently under way~\cite{paper2}.

Finally, although an analytical understanding of the linear stage of preheating is often very useful to study its
cosmological consequences, lattice simulations are eventually necessary to further study the non-linear dynamics. This is
a difficult task in general for preheating after small field inflation, because of the wide range of scales that appear
in the problem. In such cases, lattice simulations can be performed only for a limited range of parameters and the
analytical results that we derived in this paper may be used to extrapolate the results to other regions of the parameter
space.

%%%%%%%%%%%%%%%%%%%%%%%%%%%%%%%%%%%%%%%%%%%%%%%%%%%%%%%%%%%%%%%%%%%%%%%%%%%%%%%%%

\section*{Acknowledgments}

It is a pleasure to thank Maxim Khlopov, David Langlois and Jihad Mourad for useful discussions.

%%%%%%%%%%%%%%%%%%%%%%%%%%%%%%%%%%%%%%%%%%%%%%%%%%%%%%%%%%%%%%%%%%%%%%%%%%%%%%%%%%%%%%%%%%%%%%%%%%%%%%%%%%%%%%

\appendix

\section{Evolution of the Inflaton Perturbations in the Crater of the Volcano}
\label{AppHyper}

In this appendix, we solve the mode equation (\ref{modeq}) in the interval $\left[-t_m , t_m\right]$ and we match the solution to the WKB solutions (\ref{alpha}-\ref{abar}). As discussed in Section~\ref{Sec2}, we do this by looking for an approximate Schr\"{o}dinger potential for which we can find exact analytical solutions.

In order to find guidance on the form of a suitable Schr\"{o}dinger potential, let us first consider the mode equation
(\ref{modeq}) in the model (\ref{potp}) with $p = 2$~\footnote{The solution for the modes in that case has been studied
numerically in the second paper of \cite{tachyonic} in the context of preheating after hybrid inflation.}. As discussed in Section~\ref{SecBack}, this model in itself does not lead to small field inflation. However, in that case, the background evolution of the inflaton condensate can be solved analytically, and we can then find the analytical expression of the Schr\"{o}dinger potential appearing in the mode equation. Thus we start by considering the rescaled inflaton potential
\be
\tilde{V} = \frac{1}{4}\,\left(1 - \phi^2\right)^2
\ee
instead of (\ref{Vtilde}). The conservation of energy analogous  to (\ref{Etilde}) then gives
\be
\dot{\phi}^2 = \phi\,\sqrt{1 - \frac{\phi^2}{2}}
\ee
when $\phi \gg \phi_e$. This can be integrated to give $\phi = \sqrt{2} / \cosh t$ where, as in Section \ref{Sec2}, we normalize the time coordinate such that $t = 0$ when $\phi$ is maximal. The effective Schr\"{o}dinger potential in the mode equation then reads
\be
-\tilde{V}''\left(\phi(t)\right) = 1 - \frac{6}{\cosh^2(t)}
\ee
which corresponds to (\ref{Upot}) for $k_{{\mathrm{max}}} = 1$, $U_0 = 6$ and $\mu = 1$. Note that the relation
(\ref{mumag}) is satisfied for these values of the parameters.

Exact analytical solutions of the Schr\"{o}dinger equation with this potential are known, see e.g.~\cite{LL}. This potential provides also a very good approximation to $-\tilde{V}''$ in the interval $\left[-t_m , t_m\right]$ for the model (\ref{potp4}), see Fig.~\ref{Uapp}. In this approximation, the mode equation (\ref{modeq}) reads
\be
\label{ShroU}
\ddot{v}_k + \left(k^2 - k_{{\mathrm{max}}}^2 + \frac{U_0}{\cosh^2(\mu t)}\right)\,v_k = 0 \ .
\ee
It is convenient to define the parameter
\be
\label{nu}
\nu = \frac{\sqrt{k_{{\mathrm{max}}}^2 - k^2}}{\mu}
\ee
and to make the change of variable
\be
x = \frac{e^{-\mu\,t}}{e^{\mu\,t} + e^{-\mu\,t}} \ .
\ee
Note that $\nu$ is real for $k \leq k_{{\mathrm{max}}}$ and pure imaginary otherwise. In terms of the function
$x^{-\nu/2}\,\left(1-x\right)^{-\nu/2}\,v_k$, Eq.~(\ref{ShroU}) reduces to the canonical equation for the hypergeometric functions $F(a,b,c,x)$ \cite{AS} with
\bea
a &=& \nu + \frac{1}{2} + \sqrt{\frac{U_0}{\mu^2} + \frac{1}{4}} \\
\label{b}
b &=& \nu + \frac{1}{2} - \sqrt{\frac{U_0}{\mu^2} + \frac{1}{4}} \\
c &=& \nu + 1 \ .
\eea
The general solution may then be written as
\be
\label{hyper}
v_k = A_k\,x^{\nu/2}\,\left(1-x\right)^{\nu/2}\,F\left(a, b, c, 1-x\right)
+ B_k\,x^{\nu/2}\,\left(1-x\right)^{-\nu/2}\,F\left(c-b, c-a, 1 - \nu, 1-x\right)
\ee
where $A_k$ and $B_k$ are constant coefficients, to be determined by the matching conditions.

We can now derive the extra condition (\ref{mumag}) between the parameters by requiring that (\ref{hyper}) reproduces the known behavior $v_k \propto \dot{\phi}$ of the zero-mode $k = 0$. At $t = 0$, the inflaton condensate reaches its maximal value and $\dot{\phi}$ changes sign, so that (\ref{hyper}) for $k = 0$ should be an odd function of $t$, i.e. it should be anti-symmetric around $x = 1/2$. One can check that this is not possible for the second term in (\ref{hyper}), so that
$B_k = 0$ for the zero-mode. In order for the first term to be anti-symmetric around $x = 1/2$, $F(a, b, c, 1-x)$ must reduce to a first-order polynomial, which requires $a = -1$ or $b = -1$ for $k = 0$. Only the second condition is possible and, using (\ref{nu}, \ref{b}) for $k = 0$, this gives
\be
\label{bem1}
\frac{1}{2} + \frac{k_{{\mathrm{max}}}}{\mu} - \sqrt{\frac{U_0}{\mu^2} + \frac{1}{4}} = -1
\ee
which is equivalent to Eq.~(\ref{mumag}) for $\mu$. The numerical value is $\mu \simeq 2.21$.

Let us now consider the behaviour of (\ref{hyper}) in the vicinity of $t = -t_m$ and $t = t_m$, where the matching has to be done. Around $t = -t_m$, $x \rightarrow 1$ where $F(a,b,c,1-x) \rightarrow 1$. This gives
\be
\label{Asmtm}
v_k \simeq A_k\,e^{\nu \mu t} + B_k\,e^{-\nu \mu t} \;\;\;\;\;\;\; \mbox{ for } t \rightarrow -t_m \ .
\ee
Around $t = t_m$, $x \rightarrow 0$ where $F(a,b,c,1-x)$ can be expanded into a power series with coefficients that involves the Euler Gamma function \cite{AS}. This gives
\be
\label{Astm}
v_k \simeq \left(A_k\,C_k + B_k\,D_k\right)\,e^{\nu \mu t} - \left(A_k\,D_k + B_k\,\bar{C}_k\right)\,e^{-\nu \mu t}
\;\;\;\;\;\;\; \mbox{ for } t \rightarrow t_m
\ee
where the coefficients
\bea
\label{Ck}
C_k = \frac{\pi\,\Gamma(c)}{\sin\left(\pi \nu\right)\,\Gamma(1-\nu)\,\Gamma(a)\,\Gamma(b)}\\
\label{Dk}
D_k = \frac{\cos\left[(a - b) \,\pi / 2\right]}{\sin\left(\pi \nu\right)} \;\;\;\;\;\; ,  \;\;\;\;\;\;
\bar{C}_k \, C_k = D_k^2 - 1
\eea
have been simplified by using  the reflection formula for the Gamma functions.

\subsection{Transfer Matrix for $k \ll k_\mathrm{max}$}

Consider first the case $k < k_\mathrm{max}$. In that case, we have to match (\ref{Asmtm}) with (\ref{a}) at
$t = -t_m$ and (\ref{Astm}) with (\ref{abar}) at $t = t_m$. Eliminating the coefficients $A_k$ and $B_k$ gives the transfer matrix
\be
\label{abara}
\left(\begin{array}{c}
\bar{a}_k \vspace*{0.1cm}\\
\bar{b}_k
\end{array}\right) =
\left(\begin{array}{cc}
- \bar{C}_k\,e^{-X_k - 2 \nu \mu t_m} & - D_k\,e^{X_k} \vspace*{0.1cm}\\
D_k\,e^{-X_k} & C_k\,e^{X_k + 2 \nu \mu t_m}
\end{array}\right) \,
\left(\begin{array}{c}
a_k \vspace*{0.1cm}\\
b_k
\end{array}\right)
\ee
where $X_k$ is defined in Eq.~(\ref{Xk}). Note that the determinant of the transfer matrix (\ref{abara}) is equal to one.

As discussed in Section~\ref{Sec2}, the WKB approximation is strictly valid only for the modes with $k \ll k_\mathrm{max}$, which dominate the tachyonic effect. In this limit, the transfer matrix simplifies significantly. First of all, in the expression for $\bar{a}_k$, the term in $\bar{C}_k\,e^{-X_k}\,a_k$ is neglibile with respect to the term in
$D_k\,e^{X_k}\,b_k$ for these modes. Next, the condition (\ref{bem1}) means that $b \rightarrow -1$ for $k \rightarrow 0$, so that $\Gamma(b) \rightarrow \infty$ and $C_k \rightarrow 0$ for these modes, see Eq.~(\ref{Ck}). From (\ref{abara}), this implies that $|\bar{a}_k / \bar{b}_k| \sim e^{2 X_k}$ for $k \rightarrow 0$, so that the solution (\ref{abar}) decreases exponentially during the whole interval of time $\left[t_m , t_0\right]$. Indeed, this was precisely how the condition (\ref{bem1}) was obtained, by imposing that the solution $v_k(t)$ for $k = 0$, which increases exponentially in the whole interval $\left[-t_0 , -t_m\right]$, be antisymmetric around $t = 0$. Now we consider the case
$0 < k \ll k_\mathrm{max}$ and compute the first order correction in $k / k_\mathrm{max}$. In this limit
$b \simeq - 1 - k^2 / (2 \mu k_\mathrm{max})$ and we can use
\be
\Gamma(- 1 - \epsilon) = \frac{\Gamma(1 - \epsilon)}{\epsilon\,(\epsilon + 1)} \rightarrow \frac{1}{\epsilon}
\;\;\;\;\;\;\; \mbox{ for } \; \epsilon \rightarrow 0
\ee
to obtain
\be
e^{2 \nu \mu t_m} \, C_k \simeq
\frac{\pi\,\Gamma(c)\,e^{2 \nu \mu t_m}}{2 \mu\,k_\mathrm{max}\,\sin\left(\pi \nu\right)\,\Gamma(1-\nu)\,\Gamma(a)}
\,k^2 = E\,\left(\frac{k}{k_\mathrm{max}}\right)^2
\ee
where the first factor can be calculated for $k = 0$, which gives $E \simeq 2.15$. To first order in $k / k_\mathrm{max}$, the factor $D_k$ in (\ref{Dk}) can also be calculated for $k = 0$, which gives $D_k \simeq 1$ after using (\ref{bem1}).
Plugging all this into (\ref{abara}) leads to the transfer matrix (\ref{abaram}).

\subsection{Transfer Matrix for $k \gg k_\mathrm{max}$}

For $k > k_\mathrm{max}$, we can use the formula above with
\be
\label{lambda}
\nu = i\,\lambda \;\;\;\;\;\;\;\; , \;\;\;\;\;\;\;\; \lambda = \frac{\sqrt{k^2 - k_{\mathrm{max}}^2}}{\mu} \, .
\ee
As discussed in Section~\ref{Sec2}, the WKB approximation is valid for $k \gg k_{\mathrm{max}}$.
In that case, we have to match (\ref{Asmtm}) with (\ref{alpha}) at $t = -t_m$ and (\ref{Astm}) with (\ref{alphabar})~\footnote{For $k > k_\mathrm{max}$, we take $t_m$ instead of $t_k^+$ in the lower limit of the time integrals in
(\ref{alphabar}).} at $t = t_m$. Eliminating the coefficients $A_k$ and $B_k$ gives the transfer matrix between the solutions (\ref{alpha}) and (\ref{alphabar}) for each oscillation of the condensate. The coefficients
$\alpha_k^{j}$ and $\beta_k^{j}$ in (\ref{alpha}) {\it after} the $j^\mathrm{th}$ complete oscillation of the condensate
are further related to the coefficients $\bar{\alpha}_k^{j-1}$ and $\bar{\beta}_k^{j-1}$ in (\ref{alphabar}) {\it during} the $j^\mathrm{th}$ oscillation by $\alpha_k^{j} = \bar{\alpha}_k^{j-1}\,e^{-i \Theta_k}$ and
$\beta_k^{j} = \bar{\beta}_k^j\,e^{i \Theta_k}$, where $\Theta_k$ is given by (\ref{Thetak}) with $t_k^+$ replaced by $t_m$ in the lower limit of the integral. We can then relate the Bogolyubov coefficients after and before the $j^\mathrm{th}$ oscillation of the inflaton condensate as
\be
\label{albaral}
\left(\begin{array}{c}
\alpha_k^j \vspace*{0.1cm}\\
\beta_k^j
\end{array}\right) =
\left(\begin{array}{cc}
- \bar{C}_k\,e^{-2 i \Psi_k} & - D_k \vspace*{0.1cm}\\
D_k & C_k\,e^{2 i \Psi_k}
\end{array}\right) \,
\left(\begin{array}{c}
\alpha_k^{j-1} \vspace*{0.1cm}\\
\beta_k^{j-1}
\end{array}\right)
\ee
where
\be
\label{Psik}
\Psi_k = \lambda \mu t_m + \int_{t_m}^{t_0} \omega_k(t)\,dt
\ee
is the total effective phase accumulated during half a period of the condensate oscillation.
The determinant of the transfer matrix (\ref{albaral}) is again equal to one.

The occupation numbers after and before the $j^\mathrm{th}$ oscillation of condensate, $n_k^j = |\beta_k^j|^2$ and
$n_k^{j-1} = |\beta_k^{j-1}|^2$, can then be related by using (\ref{albaral}) and the normalization condition
$|\alpha_k^{j-1}|^2 - |\beta_k^{j-1}|^2 = 1$. This gives
\be
\label{nkCD}
n_k^j = |D_k|^2 + \left(|C_k|^2 + |D_k|^2\right)\,n_k^{j-1} + 2\,|C_k|\,|D_k|\,\sqrt{n_k^{j-1}\,(n_k^{j-1} + 1)}\,
\cos \theta_k^j
\ee
where $\theta_k^j = 2 \Psi_k + \arg C_k - \arg D_k + \arg \beta_k^{j-1} - \arg \alpha_k^{j-1}$. To calculate $|D_k|^2$,
we can use (\ref{bem1}) and (\ref{lambda}) in (\ref{Dk}) to obtain (\ref{Dk2}). Finally, using (\ref{bem1}) and
(\ref{lambda}) in (\ref{Ck}), we can simplify $|C_k|^2$ by using the reflection formula for the Gamma function and the fact that the complex conjugate of $\Gamma(z)$ is given by $\Gamma(z^*)$. This gives $|C_k|^2 = 1 + |D_k|^2$ and
Eq.~(\ref{nkCD}) then reduces to Eq.~(\ref{nkD}).

%%%%%%%%%%%%%%%%%%%%%%%%%%%%%%%%%%%%%%%%%%%%%%%%%%%%%%%%%%%%%

%%%%%%%%%%%%%%%%%%%%%%%%%%%%%%%%%%%%%%%%%%%%%%%%%%%%%%%%%%%%%%%%%%%%%%%%%

\end{document}